\documentclass[11pt]{article}

\RequirePackage[authoryear]{natbib}

\usepackage{color}
\usepackage{url}
\usepackage[margin=1in]{geometry}
\usepackage{amsmath,amssymb}
\usepackage{verbatim}
\usepackage{graphicx,setspace}
\usepackage{amsthm}
\usepackage{subcaption}

\usepackage{tikz}
\usetikzlibrary{bayesnet}
\usepackage{import}
\newtheorem{theorem}{\noindent Theorem}

\def\bfmath#1{\mathchoice
        {\mbox{\boldmath$#1$}}%
        {\mbox{\boldmath$#1$}}%
        {\mbox{\boldmath$\scriptstyle#1$}}%
        {\mbox{\boldmath$\scriptscriptstyle#1$}}}%

\def\timeInd{d}
\def\pieces{D}

\def\Rate{\mathbf{\Lambda}}
\def\proposal{Q}
\def\isRate{q}
\def\isQ{\phi}
\def\isQbold{\bfmath{\phi}}
\def\stationaryLambdaTilde#1{\tilde{\mbox{\boldmath$\lambda$}}\mbox{$^{#1}$}}
\def\stationaryGamma{\bfmath{\gamma}}
\def\stationaryLambda{\bfmath{\lambda}}
\def\bfA{\mathbf{A}}
\def\bfP{\mathbf{P}}
\def\bfb{\mathbf{b}}
\def\bfv{\mathbf{v}}
\def\moran{\mathbf{M}}
\def\bfGamma{\mathbf{\Gamma}}
\def\moranModified{\mathbf{\tilde{M}}}

\def\Lconst{\mathcal{L}_\text{const}}
\def\Lexact{\mathcal{L}_\text{exact}}
\def\Lapprox{\mathcal{L}_\text{approx}}
\def\Lmiss{\mathcal{L}_\text{miss}}

\def\bfp{\bfmath{p}}
\def\bfn{\mathbf{n}}
\begin{document}

\begin{center}
\begin{spacing}{1.5}
\textbf{\Large  Two-Locus Likelihoods under Variable Population Size and Fine-Scale Recombination Rate Estimation}
\end{spacing}

\vspace{5mm}
John A. Kamm$^{1,2,*}$, Jeffrey P. Spence$^{3,*}$, Jeffrey Chan$^2$, and Yun S. Song$^{1,2,4,5}$

\vspace{8mm}
$^1$ Department of Statistics, University of California, Berkeley, CA 94720\\
$^2$ Computer Science Division, University of California, Berkeley, CA 94720\\
$^3$ Computational Biology Graduate Group, University of California, Berkeley, CA 94720\\
$^4$ Department of Integrative Biology, University of California, Berkeley, CA 94720\\
$^5$ Department of Mathematics and Department of Biology, University of Pennsylvania, PA~19104\\
$^*$ These authors contributed equally to this work.
\end{center}

\begin{abstract}
Two-locus sampling probabilities have played a central role in devising an efficient composite likelihood method for estimating fine-scale recombination rates.  Due to mathematical and computational challenges, these sampling probabilities are typically computed under the unrealistic assumption of a constant population size, and simulation studies have shown that resulting recombination rate estimates can be severely biased in certain cases of historical population size changes.  To alleviate this problem, we develop here new methods to compute the sampling probability for variable population size functions that are piecewise constant.  Our main theoretical result, implemented in a new software package called LDpop, is a novel formula for the sampling probability that can be evaluated by numerically exponentiating a large but sparse matrix.  This formula can handle moderate sample sizes ($n \leq 50$) and demographic size histories with a large number of epochs ($\mathcal{D} \geq 64$).  In addition, LDpop implements an approximate formula for the sampling probability that is reasonably accurate and scales to hundreds in sample size ($n \geq 256$).  Finally, LDpop includes an importance sampler for the posterior distribution of two-locus genealogies, based on a new result for the optimal proposal distribution in the variable-size setting.  Using our methods, we study how a sharp population bottleneck followed by rapid growth affects the correlation between partially linked sites.  Then, through an extensive simulation study, we show that accounting for population size changes under such a demographic model leads to substantial improvements in fine-scale recombination rate estimation. 
LDpop is freely available for download at \url{https://github.com/popgenmethods/ldpop}.
\end{abstract}

\section{Introduction}

The coalescent with recombination \citep{griffiths1997ancestral} provides a basic population genetic model for
recombination.
For a very small number of loci and a constant population size,
the likelihood (or sampling probability) can be computed via a
recursion \citep{golding1984sampling, ethier1990two, hudson2001two} or importance
sampling \citep{fearnhead2001estimating}, allowing for
maximum-likelihood and Bayesian estimates of recombination rates
\citep{fearnhead2001estimating,hudson2001two,mcvean2002coalescent,fearnhead2004application,fearnhead2005novel,fearnhead2006sequenceldhot}.

\citet{jenkins2009closed,jenkins2010asymptotic} recently introduced a new framework based on asymptotic series (in inverse powers of the recombination rate $\rho$) to approximate the two-locus sampling probability under a constant population size, and 
developed an algorithm for finding the expansion to an arbitrary order \citep{jenkins2012pade}.  They also proved that only a finite number of terms in the expansion is needed to obtain the \emph{exact} two-locus sampling probability as an analytic function of $\rho$.  
\citet{bhaskar2012closed} partially extended this approach to an arbitrary number of loci and found closed-form formulas for the first two terms in an asymptotic expansion of the multi-locus sampling distribution.

When there are more than a handful of loci,
computing the sampling probability becomes intractable.
A popular and tractable alternative has been to construct composite likelihoods by multiplying the two-locus likelihoods for pairs of SNPs;
this pairwise composite likelihood has been used to estimate fine-scale recombination rates in humans
\citep{ihmc2007seconda, gpc2010map},
\emph{Drosophila }\citep{chan2012genome}, chimpanzees \citep{auton2012fine},
microbes \citep{johnson2009inference}, dogs \citep{auton2013genetic}, and more, and
was used in the discovery of a DNA motif associated with recombination
hotspots in some organisms, including humans \citep{myers2008common}, subsequently identified as a binding site of the
protein PRDM9 \citep{myers2010drive, baudat2010prdm9, berg2010prdm9}.

The pairwise composite likelihood was first suggested by
\citet{hudson2001two}.
The software package LDhat \citep{mcvean2004fine, auton2007recombination}
implemented the pairwise composite likelihood and embedded it within a
Bayesian MCMC algorithm for inference.
\citet{chan2012genome} modified this algorithm
in their program LDhelmet to efficiently utilize aforementioned asymptotic
formulas for the sampling probability,
among other improvements.
The program LDhot \citep{myers2005fine, auton2014identifying}
uses the composite likelihood
as a test statistic to detect recombination hotspots, in conjunction with
coalescent simulation to determine appropriate null distributions.

Because of mathematical and computational challenges, 
LDhat, LDhelmet, and LDhot all assume a constant population size model to compute the two-locus sampling probabilities.
This is an unrealistic assumption, and it would be desirable to account for known demographic events,
such as bottlenecks or population growth.
Previous studies \citep{mcvean2002coalescent,chan2012genome,smith2005comparison} have shown
that incorrectly assuming constant population size can lead these composite-likelihood methods
to produce biased estimates.
Furthermore, \citet{johnston2012population} observed that a sharp bottleneck followed by rapid growth
can lead LDhat to infer spurious recombination hotspots if it assumes a constant population size.

\citet{hudson2001two} proposed Monte Carlo computation of two locus likelihoods by simulating genealogies.
While this generalizes to arbitrarily complex demographies, it would be desirable to have a deterministic formula, as naive Monte Carlo computation sometimes has difficulty approximating small probabilities.

In this paper, we show how to compute the two-locus sampling probability exactly 
under variable population size histories that are piecewise constant.
Our approach relies on the Moran model \citep{moran1958random, ethier1993fleming, donnelly1999genealogical},
a stochastic process closely related to the coalescent.
We have implemented our results in a freely available software package, LDpop,
that efficiently produces lookup tables of two-locus likelihoods under variable population size.
These lookup tables can then be used by other programs that use composite likelihoods to infer recombination maps.

Our main result is an exact formula,
introduced in Theorem~\ref{thm:augmented:moran}, that involves
exponentiating sparse $m$-by-$m$ matrices
containing $O(m)$ nonzero entries, where $m = O(n^6)$ under a bi-allelic model, with $n$ being the sample size.
We derive this formula by constructing a Moran-like process in which sample paths can be coupled with the two-locus coalescent, and by applying a reversibility argument.

Theorem~\ref{thm:augmented:moran} has a high computational cost, and our implementation in LDpop can practically handle low to moderate sample sizes ($n < 50$) on a 24-core compute server.
We have thus implemented an approximate formula that is much faster, and scales to sample sizes in the hundreds.
This formula is computed by exponentiating a sparse matrix with $O(n^3)$ nonzero entries,
and is based on a previous two-locus Moran process \citep{ethier1993fleming},
which we have implemented and extended to the case of variable population size.
While this formula does not give the exact likelihood, it provides a reasonable approximation,
and converges to the true value in an appropriate limit.

In addition to these exact and approximate formulas, LDpop also includes a
highly efficient importance sampler for the posterior distribution
of two-locus genealogies. This can be used to infer the genealogy
at a pair of sites, and also provides an alternative method for computing
two-locus likelihoods.
Our importance sampler is based on an optimal proposal distribution that we characterize
in Theorem~\ref{thm:posterior:chain}.
It generalizes previous results
for the constant size case, which have been used
to construct importance samplers for both the single-population, two-locus case \citep{fearnhead2001estimating, dialdestoro2016coalescent}
and for other constant-demography coalescent scenarios \citep{stephens2000inference,deiorio2004importance,griffiths2008importance,hobolth2008importance,jenkins2012stopping,koskela2015computational}.
The key ideas of Theorem~\ref{thm:posterior:chain}
should similarly generalize to other contexts of
importance sampling a time-inhomogeneous coalescent.

Using a simulation study, we show that using LDpop to account
for demography substantially improves the composite-likelihood inference
of recombination maps.
We also use LDpop to gain a qualitative understanding of linkage disequilibrium by examining the $r^2$ statistic.
Finally, we examine how LDpop scales in terms of sample size $n$ and the number $\mathcal{D}$ of
demographic epochs.
The exact formula can handle $n$ in the tens, while the approximate formula can handle
$n$ in the hundreds.
Additionally, we find that the runtime of LDpop is not very sensitive to $\mathcal{D}$,
so LDpop can handle size histories with a large number of pieces.

\bigskip
\emph{Software availability:}
Our software package LDpop is freely available for download at\break \url{https://github.com/popgenmethods/ldpop}.

\section{Background}
Here we describe our notational convention, and review some key concepts regarding the coalescent with recombination and the two-locus Moran model.

\subsection{Notation}
Let $\frac{\theta}2$ denote the mutation rate
per locus per unit time, $\bfP=(P_{ij})_{i,j\in\mathcal{A}}$ the transition probabilities
between alleles given a mutation, and $\mathcal{A} = \{0,1\}$ the set of alleles
(our formulas can be generalized to $|\mathcal{A}| > 2$, but this increases the computational complexity).
Let $\frac{\rho}2$ denote the recombination rate
per unit time.
We consider a single panmictic population,
with piecewise-constant effective population sizes.
In particular, we assume $\pieces$ pieces,
with endpoints $-\infty = t_{-\pieces} < t_{-\pieces+1} < \cdots < t_{-1} < t_0 = 0$, where $0$ corresponds to the present and  $t< 0$ corresponds to a time in the past.
The piece $(t_\timeInd, t_{\timeInd+1}]$ is assumed to have scaled population size $\eta_{\timeInd}$.
Going backwards in time, two lineages coalesce (find a common ancestor)
at rate $\frac1{\eta_{\timeInd}}$ within the interval $(t_{\timeInd}, t_{\timeInd+1}]$.

We allow the haplotypes to have missing (unobserved)
alleles at each locus, and use $*$ to denote such alleles.
We denote each haplotype as having type $a$, $b$, or $c$,
where $a$ haplotypes are only observed at the left locus,
$b$ haplotypes are only observed at the right locus,
and $c$ haplotypes are observed at both loci.
Overloading notation, we sometimes refer to the left locus as the $a$ locus,
and the right locus as the $b$ locus.
We use $\mathbf{n}=\{n_{i*},n_{*j}, n_{k\ell}\}_{i,j,k,\ell\in\mathcal{A}}$ to denote the configuration of an unordered collection of two-locus haplotypes, with $n_{i*}$ denoting the number of $a$-types with allele $i$, $n_{*j}$ denoting the number of $b$-types with allele $j$, and $n_{k\ell}$ denoting the number of $c$-types with alleles $k$ and $\ell$.

Suppose $\mathbf{n}$ has $n^{(abc)} = (n^{(a)},n^{(b)},n^{(c)})$ haplotypes
of type $a,b,c$ respectively. 
We define the \emph{sampling probability} $\mathbb{P}_t(\mathbf{n})$
to be the probability of sampling $\mathbf{n}$ at time $t$,
given that we observed $n^{(a)},n^{(b)},n^{(c)}$ haplotypes of type
$a,b,c$, under the coalescent with recombination
(next subsection).

\subsection{The ARG and the coalescent with recombination \label{sec:coal:recom}}

 \begin{figure}[t]
\centering
\def\svgwidth{.7\textwidth}
   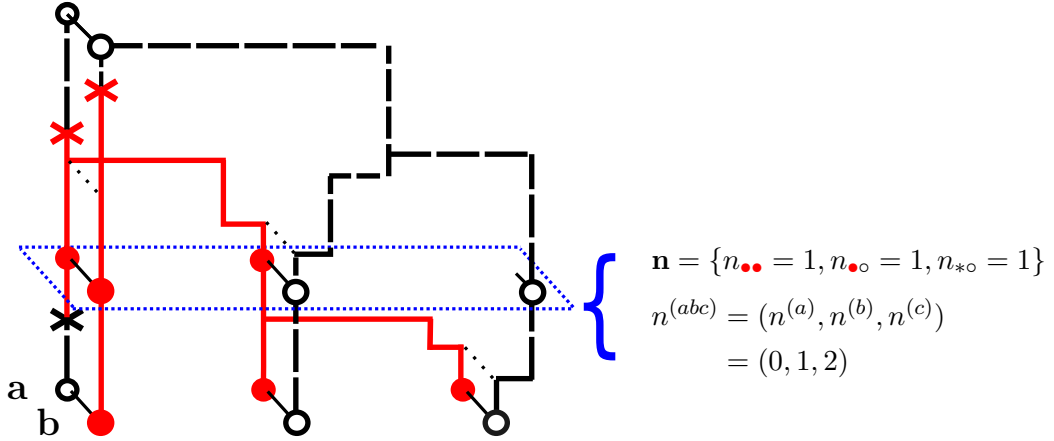
  \caption{An ancestral recombination graph (ARG) at two loci, labeled $a$ and $b$, each with 
    two alleles (${\color{red} \bullet}$ and $\circ$).
The notation $\mathbf{n}, n_{ij}, n^{(abc)}$ are illustrated for the configuration between the first
coalescence and second recombination event.}
  \label{fig:arg}
\end{figure}

The Ancestral Recombination Graph (ARG) is
the multi-locus genealogy relating a sample (Figure~\ref{fig:arg}).
The coalescent with recombination \citep{griffiths1991two} gives the limiting
distribution of the ARG under a wide class of population models,
including the Wright-Fisher model and the Moran model.

Let $n^{(c)}_t$ be the number of lineages at time $t$
that are ancestral to the observed present-day sample
at both loci. Similarly, let $n^{(a)}_t$ and $n^{(b)}_t$
be the number of lineages that are ancestral at only the
$a$ or $b$ locus, respectively.
Under the coalescent with recombination,
$n^{(abc)}_t = (n_t^{(a)},n_t^{(b)},n_t^{(c)})$
is a backward in time Markov chain, 
where each $c$ type lineage splits (recombines)
into one $a$ and one $b$ lineage at rate $\frac{\rho}2$,
and each pair of lineages coalesces at rate $\frac{1}{\eta_\timeInd}$ within the time interval $(t_\timeInd, t_{\timeInd+1}]$.
Table~\ref{tab:arg:rates} gives the transition rates of $n^{(abc)}_t$.

\begin{table}[t]
  \caption{Backward in time transition rates of
$n^{(abc)}_t = (n_t^{(a)},n_t^{(b)},n_t^{(c)})$ within time interval $(t_\timeInd, t_{\timeInd+1}]$ under the coalescent with recombination.}
  \label{tab:arg:rates}
  \centering
  \begin{tabular}{cc}
\hline
End state & Rate 
\\ \hline
$ n_t^{(a)}+1, n_t^{(b)}+1, n_t^{(c)} - 1$ & $\frac{\rho}2 n_t^{(c)}$ \\ 
$ n_t^{(a)}-1,  n_t^{(b)}, n_t^{(c)}$ & $\frac1{\eta_\timeInd} n_t^{(a)} \big(\frac{n_t^{(a)}-1}2 +  n_t^{(c)}\big)$ \\ 
$ n_t^{(a)},  n_t^{(b)}-1, n_t^{(c)}$ & $\frac1{\eta_\timeInd} n_t^{(b)} \big(\frac{n_t^{(b)}-1}2 +  n_t^{(c)}\big)$ \\ 
$ n_t^{(a)},  n_t^{(b)}, n_t^{(c)}-1$ & $\frac1{\eta_\timeInd} {n_t^{(c)} \choose 2}$\\ 
$ n_t^{(a)}-1,  n_t^{(b)}-1, n_t^{(c)}+1$ &$\frac1{\eta_\timeInd} n_t^{(a)} n_t^{(b)}$ \\ \hline
  \end{tabular}
\end{table}

After sampling the history of coalescence and recombination events
$\{n^{(abc)}_t\}_{t \leq 0}$,
we drop mutations down at rate $\frac{\theta}2$ per locus, with
alleles mutating according to $\bfP$, and the alleles of the common
ancestor assumed to be at the stationary distribution.
This gives us a sample path $\{\mathbf{n}_t\}_{t \leq 0}$,
where $\mathbf{n}_0$ is the observed sample at the present,
and $\mathbf{n}_t$ is the collection of ancestral haplotypes at time $t$.
Under this notation, the sampling probability at time $t$ is defined as
\begin{eqnarray}
  \mathbb{P}_t(\mathbf{n})
&:=&
\mathbb{P} (\mathbf{n}_t = \mathbf{n} \mid n^{(abc)}_t = n^{(abc)}).
\label{eq:P_ARG}
\end{eqnarray}

\subsection{Two-locus Moran model \label{sec:moran:dk}}

The Moran model is a stochastic process
closely related to the coalescent, and plays
a central role in our results.
Here, we
review a two-locus Moran model with recombination dynamics from
\citet{ethier1993fleming}.
We note that there are multiple versions of the two-locus Moran model,
and in particular \citet{donnelly1999genealogical} describe a Moran model
with different recombination dynamics.

The Moran model with $N$ lineages is a
finite population model evolving forward in time.
In particular, let $\moran_t$ denote a collection
of $N$ two-locus haplotypes at time $t$ (with no missing alleles).
Then $\moran_t$ is a Markov chain
going forwards in time that changes due
to mutation, recombination, and copying events.

\begin{figure}[t]
  \centering
  \includegraphics[width=.4\textwidth]{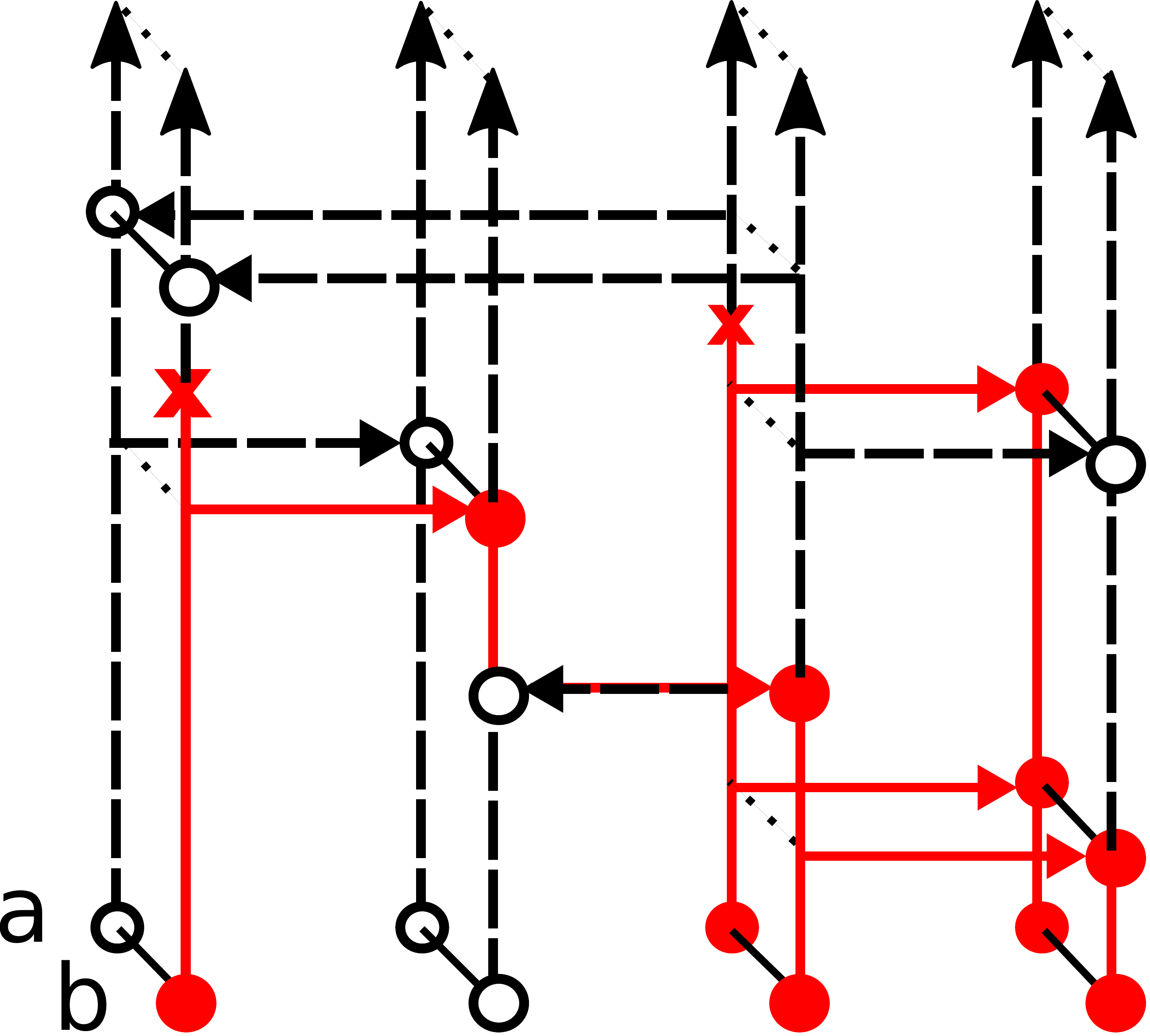}
  \caption{A finite two-locus Moran model with $N=4$ particles. Each lineage copies itself onto every other lineage at rate $\frac{1}{2 \eta}$. Mutations arise at rate $\frac{\theta}2$ per allele per locus. Recombination follows dynamics from \citet{ethier1993fleming}: every pair of lineages experience a crossover recombination at rate $\frac{\rho}{2 (N-1)}$. Here, the 2nd and 3rd lineages swap their $b$ alleles through a crossover. The sampling probability for this model agrees with the coalescent at each locus marginally, but not jointly at both loci (though the discrepancy disappears as $N \to \infty$).}
  \label{fig:moran}
\end{figure}

Let $\Rate^{\timeInd}_{(N)}$ denote the transition matrix of $\moran_t$
within $(t_\timeInd, t_{\timeInd+1}]$.
We describe the rates of $\Rate^{\timeInd}_{(N)}$.
For the mutation events, each allele mutates
at rate $\frac{\theta}2$ according to transition matrix $\bfP$.
For the copying events,
each lineage of $\moran_t$
copies its haplotype onto each other lineage
at rate $\frac1{2 \eta_\timeInd}$ within the time interval
$(t_\timeInd, t_{\timeInd+1}]$.
Biologically, this corresponds to one lineage dying out,
and being replaced by the offspring of another lineage,
which occurs more frequently when genetic drift is high
(i.e. when the population size $\eta_\timeInd$ is small).
Finally, every pair of lineages in $\moran_t$ swap genetic
material through a crossover recombination at
rate $\frac{\rho}{2(N-1)}$.
A crossover between haplotypes $(i_1,j_1)$ and $(i_2,j_2)$ results in new haplotypes
$(i_1,j_2)$ and $(i_2,j_1)$, and the configuration resulting from the crossover
is $\moran_t - \mathbf{e}_{i_1,j_1} - \mathbf{e}_{i_2,j_2} + \mathbf{e}_{i_1,j_2} + \mathbf{e}_{i_2,j_1}$.
See Figure~\ref{fig:moran} for illustration.

Let $\mathbb{P}_t^{(N)}(\mathbf{n})$ be the probability of sampling $\mathbf{n}$ at time $t$ under this
Moran model, for a configuration $\mathbf{n}$ with sample size $n \leq N$.
$\mathbb{P}_t^{(N)}(\mathbf{n})$ is given by first sampling $\moran_{\min(t , t_{-\pieces+1})}$ from the stationary
distribution $\stationaryLambda^{-\pieces}_{(N)}$ of $\Rate^{-\pieces}_{(N)}$,
then propagating $\{\moran_s\}_{s \leq 0}$ forward-in-time to $t$, and then sampling
$\mathbf{n}$ without replacement from $\moran_t$. So,
\begin{eqnarray}
  [\mathbb{P}_{(N)}(\moran_t = \moran)]_{\moran}
&=&
\stationaryLambda^{-\pieces}_{(N)} 
\prod_{\timeInd = -\pieces+1}^{-1}
e^{\Rate_{(N)}^\timeInd [\min(t,t_{\timeInd+1}) - \min(t,t_\timeInd)]} \notag \\
  \mathbb{P}_t^{(N)}(\mathbf{n})
&=&
\sum_{\moran} \mathbb{P}_{(N)}(\moran_t = \moran)
\mathbb{P}(\mathbf{n} \mid \moran),
\label{eq:moran}
\end{eqnarray}
where $\mathbb{P}(\mathbf{n} \mid \moran)$ denotes
the probability of sampling $\mathbf{n}$ without replacement from $\moran$, and
$[\mathbb{P}_{(N)}(\moran_t = \moran)]_{\moran}$ and $\stationaryLambda^{-\pieces}_{(N)}$ are row vectors here.

In general, $\mathbb{P}^{(N)}_t(\mathbf{n}) \neq \mathbb{P}_t(\mathbf{n})$, so
$\moran_t$ disagrees with the coalescent with recombination.
However, the likelihood under $\moran_t$ converges to the correct value, $\mathbb{P}^{(N)}_t(\mathbf{n}) \to \mathbb{P}_t(\mathbf{n})$ as $N\to \infty$.
In fact, even for $N=n=20$, we find that $\mathbb{P}^{(N)}_t(\mathbf{n})$ provides a reasonable
approximation for practical purposes (Sections~\ref{sec:ldhat}, \ref{sec:approx:accuracy}).
We refer to \eqref{eq:moran}, i.e. the likelihood under $\moran_t$, as the ``approximate likelihood formula'',
in contrast to the exact formula we present in Theorem~\ref{thm:augmented:moran} below.
This approximate formula is included in LDpop as a faster, more scalable alternative
to the exact formula of Theorem~\ref{thm:augmented:moran}.

\section{Theoretical Results \label{sec:results}}

In this section, we describe our theoretical results. 
Proofs are deferred to Appendix~\ref{sec:proofs}.

\subsection{Exact formula for the sampling probability \label{sec:sampling:formula}}

Our main result is an explicit formula for the sampling probability $\mathbb{P}(\mathbf{n})$, presented in Theorem~\ref{thm:augmented:moran}.
We present an outline here; the proof is delayed to Appendix~\ref{sec:proof:augmented:moran}.

The idea is to construct a forward-in-time Markov process $\moranModified_t$, and relate its distribution to the coalescent
with recombination.
$\moranModified_t$ is similar to the Moran model $\moran_t$ in Section~\ref{sec:moran:dk},
except that $\moranModified_t$ allows partially specified $a$- and $b$-types, whereas all lineages in $\moran_t$ are fully specified $c$-types.
Specifically, the state space of $\moranModified_t$ is 
$\mathcal{N} = \{\mathbf{n} : n^{(abc)} = (k,k,n-k), 0 \leq k \leq n\}$
the collection of sample configurations with $n$ specified (non-missing) alleles  at each locus.
The state of $\moranModified_t$ changes due to copying, mutation, recombination, and ``recoalescence''
events.
Copying and mutation dynamics are similar to $\moran_t$, but recombination is different:
every $c$-type splits into $a$- and $b$-types at rate $\frac{\rho}2$, and
every pair of $a$- and $b$-types ``recoalesces'' back into a $c$-type at rate $\frac1{\eta_\timeInd}$.
An illustration
is shown in Figure~\ref{fig:augmented:moran:2};  $\moranModified_t$ is described in more detail
in Appendix~\ref{sec:proof:augmented:moran}.
Within interval $(t_{\timeInd}, t_{\timeInd+1}]$, we denote the transition rate matrix of
$\moranModified_t$ as $\tilde{\Rate}^d$, a square matrix indexed by $\mathcal{N}$,
with entries given in Table~\ref{tab:augmented:rates}.

\begin{figure}[t]
    \centering
    \begin{subfigure}[t]{0.43\textwidth}
\includegraphics[width=\textwidth]{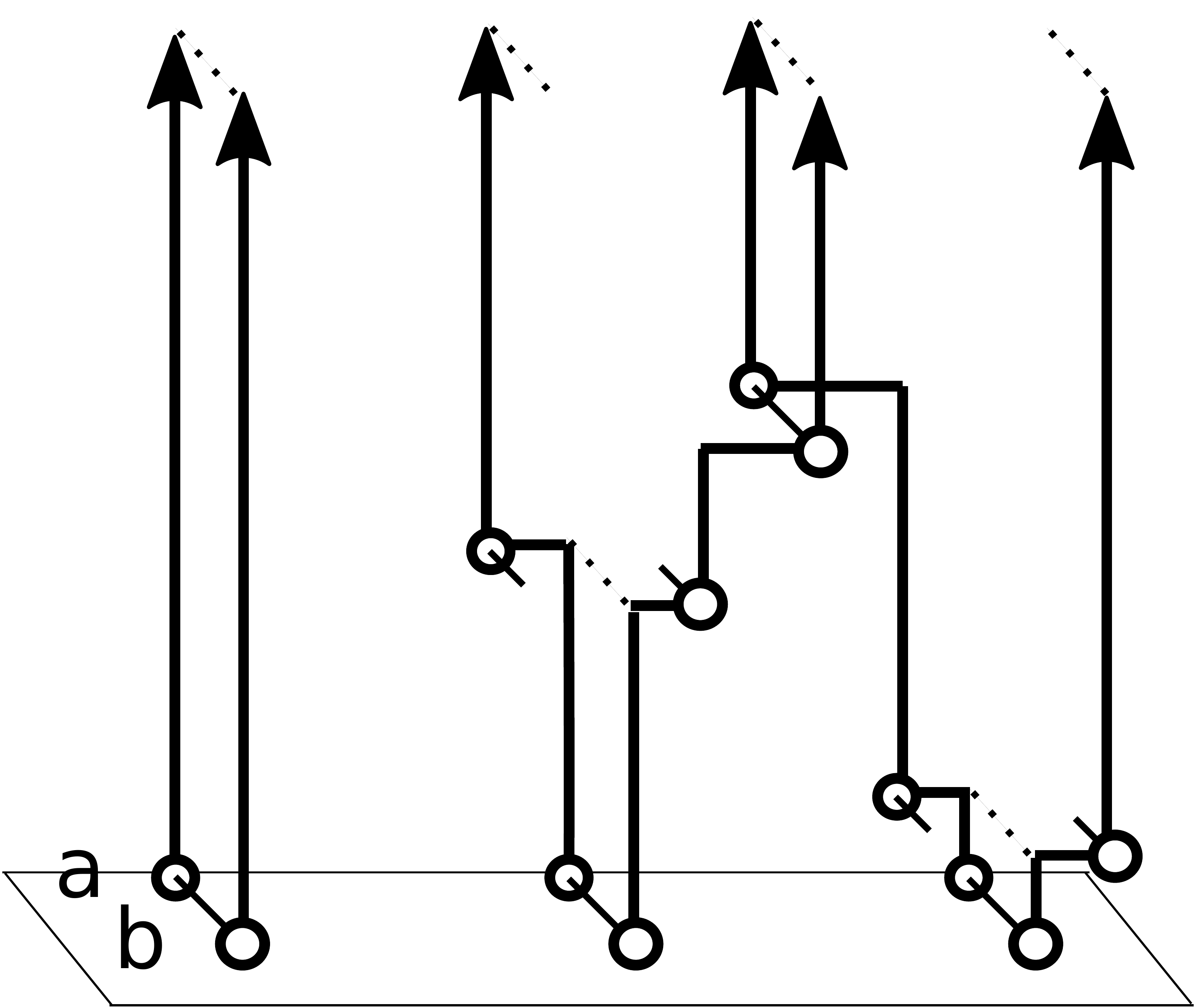}
        \caption{}
        \label{fig:augmented:moran:1}
    \end{subfigure}
    \qquad \qquad 
    \begin{subfigure}[t]{0.43\textwidth}
\includegraphics[width=\textwidth]{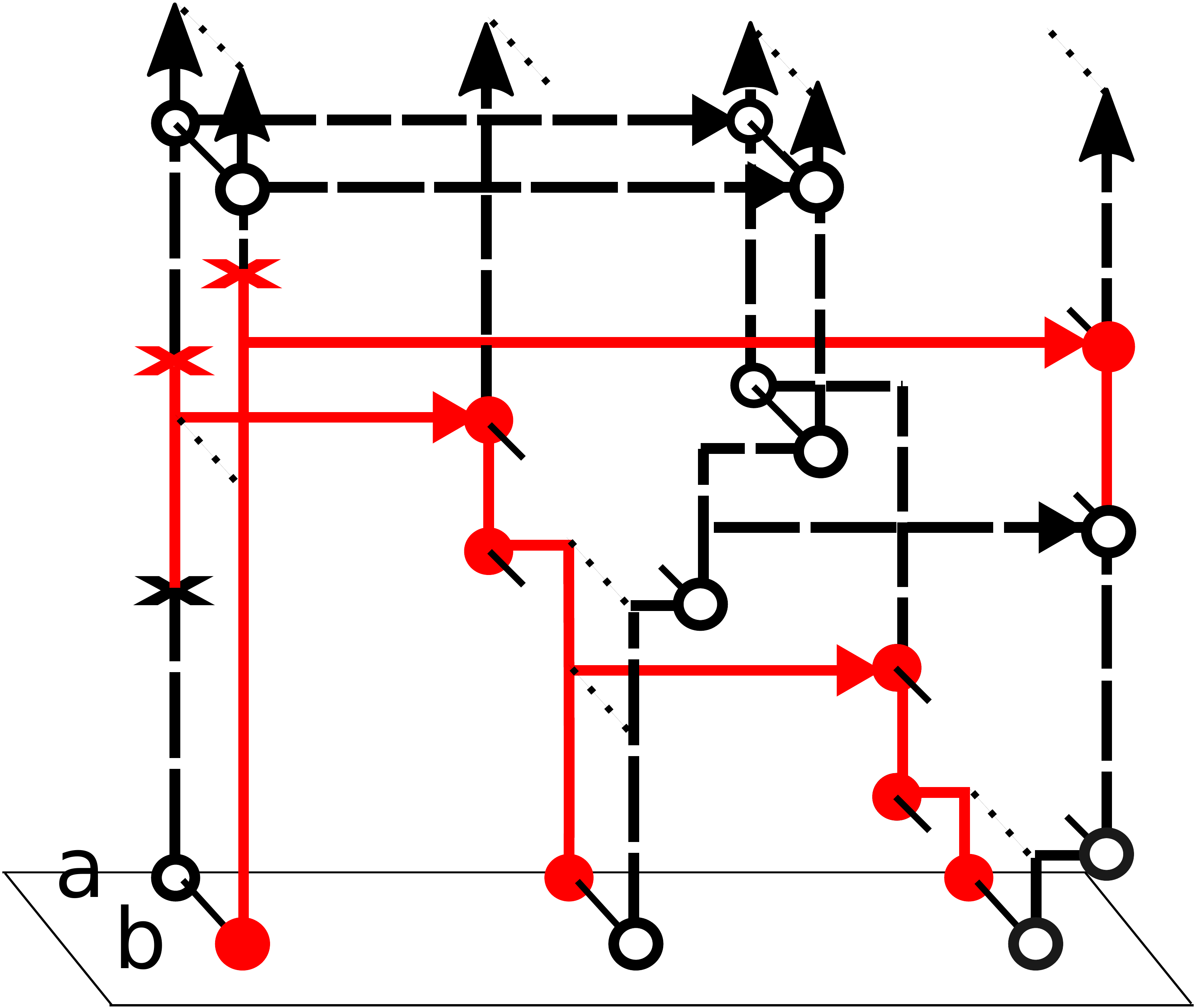}
        \caption{}
        \label{fig:augmented:moran:2}
    \end{subfigure}
    \caption{The process $\{\moranModified_t\}_{t \leq 0}$ with rates $\tilde{\Rate}^d$ (Table~\ref{tab:augmented:rates}) used to prove Theorem~\ref{thm:augmented:moran}. This process  is similar to $\{\moran_t\}$ (Figure~\ref{fig:moran}), in that $n$ is fixed, and alleles change due to copying and mutation. However $\moranModified_t$ allows partially specified $a$- and $b$-types, with $c$-types recombining into a pair of $a$- and $b$-types, and pairs of $a$- and $b$-types ``recoalescing'' into $c$-types.
    (a) The process $\{C_t\}_{t \leq 0}$ of just recombination ($c \to a,b$) and ``recoalescence'' ($a,b \to c$) events.
    (b) The full process $\{\moranModified_t\}_{t \leq 0}$ including copying, mutation, recombination, and ``recoalescence'' events.
    }\label{fig:augmented:moran}
\end{figure}

\begin{table}[t]
  \caption{Nonzero entries of the rate matrix $\tilde{\Rate}^d$ for the interval $(t_d,t_{d+1}]$.}
  \label{tab:augmented:rates}
  \centering
  \begin{tabular}[h]{cc}
    \hline
    $\mathbf{m}$ & $\tilde{\Rate}^d_{\mathbf{n},\mathbf{m}}$
\\ \hline
$\mathbf{n} - \mathbf{e}_{i*} + \mathbf{e}_{j*}$
&
$\frac1{\eta_\timeInd} n_{i*} (\frac{1}2 n_{j*} + \sum_{k\in\mathcal{A}} n_{jk})
  + \frac{\theta}2 P_{ij} n_{i*}$
\\ 
$\mathbf{n} - \mathbf{e}_{*i} +
    \mathbf{e}_{*j}$
&
$\frac1{\eta_\timeInd} n_{*i} (\frac{1}2  n_{*j} + \sum_{k\in\mathcal{A}}
n_{kj})
  + \frac{\theta}2 P_{ij} n_{*i}$
\\ 
$\mathbf{n} - \mathbf{e}_{ij} +
    \mathbf{e}_{kl}$
&
$\frac1{2\eta_\timeInd} n_{ij} n_{kl}
+ \frac{\theta}2 (\delta_{ik} P_{jl} + \delta_{jl} P_{ik}) n_{ij}$
\\ 
$\mathbf{n} - \mathbf{e}_{ij} +
    \mathbf{e}_{i*} + \mathbf{e}_{*j}$
&
$\frac{\rho}2 n_{ij}$
\\
$\mathbf{n} - \mathbf{e}_{i*} - \mathbf{e}_{*j} + \mathbf{e}_{ij}$
&
$n_{i*}n_{*j} \frac1{\eta_\timeInd}$
\\
$\mathbf{n}$
&
$- \frac1{\eta_d} {n^{(a)}+n^{(b)}+n^{(c)} \choose 2}
- \frac{\rho}2 n^{(c)}
- \frac{\theta}2 \sum_{i \in \mathcal{A}} \sum_{j \in \mathcal{A} \cup \{*\}} (n_{ij} + n_{ji})$
\\ \hline
  \end{tabular}
\end{table}

$\moranModified_t$ does not itself yield the correct sampling probability.
The basic issue is that $\moranModified_t$ has $c$-types splitting into $a$- and $b$-types
at rate $\frac{\rho}2$ going \emph{forwards in time}, but under the coalescent with recombination,
this needs to happen at rate $\frac{\rho}2$ going \emph{backwards in time}.
Similarly, pairs of $a$- and $b$-types merge into a single $c$-type at rate $\frac1{\eta_d}$
going forwards in time under $\moranModified_t$, but going backwards in time under the coalescent with recombination.

However, it is possible to ``reverse'' the direction of the recombinations ($c \to a,b$) and recoalescences ($a,b\to c$),
to get a new process that does match the two-locus coalescent.
In particular, let $C_t$ be the number of $c$ types in $\moranModified_t$, illustrated
in Figure~\ref{fig:augmented:moran:1}.
Then $C_t$ is a reversible Markov chain, whose rate matrix in $(t_\timeInd, t_{\timeInd+1}]$
is $\bfGamma^d$ a tridiagonal square matrix indexed by $\{0,1,\ldots,n\}$,
with $\Gamma^d_{m,m-1} = \frac{\rho}2 m$, $\Gamma^d_{m,m+1} = (n-m)^2 \frac1{\eta_\timeInd}$, and $\Gamma^d_{m,m} = - \Gamma^d_{m,m-1} - \Gamma^d_{m,m+1}$.
Exploiting the reversibility of $C_t$ allows us to relate the distribution of $\moranModified_t$
to the coalescent with recombination, and obtain the following result.

\begin{theorem}\label{thm:augmented:moran}
Let $(\gamma^\timeInd_0,\ldots,\gamma^\timeInd_n)$ be the stationary distribution
of $\bfGamma^\timeInd$, and let the row vector $\tilde{\stationaryGamma}^{\timeInd}$ be indexed by $\mathcal{N}$,
with $\tilde{\gamma}^{\timeInd}_{\mathbf{n}} = \gamma^{\timeInd}_m$
if $\mathbf{n}$ has $m$ lineages of type $c$.
Denote the stationary distribution of $\tilde{\Rate}^\timeInd$ by the row vector $\stationaryLambdaTilde{\timeInd}=(\lambda^\timeInd_\bfn)_{\bfn\in\mathcal{N}}$.
Let $\odot$ and $\div$ denote component-wise multiplication and division, and 
recursively define the row vector $\bfp^{\timeInd}=(p^\timeInd_{\mathbf{n}})_{\mathbf{n}\in\mathcal{N}}$ by
\begin{align}
  \bfp^{-\pieces+1} &= \stationaryLambdaTilde{-\pieces} \div \tilde{\stationaryGamma}^{-\pieces} \nonumber
\\
\bfp^{\timeInd+1} &= \big[(\bfp^{\timeInd} \odot \tilde{\stationaryGamma}^{\timeInd})
e^{\tilde{\Rate}^\timeInd (t_{\timeInd+1} - t_\timeInd)}\big] \div \tilde{\stationaryGamma}^{\timeInd}.
\label{eq:thm1}
\end{align}
Then, for $\mathbf{n} \in \mathcal{N}$, we have $\mathbb{P}_0(\mathbf{n}) = p^{0}_{\mathbf{n}}$.
\end{theorem}

Note that Theorem~\ref{thm:augmented:moran} gives
$\mathbb{P}_0(\mathbf{n})$ for $\mathbf{n} \in \mathcal{N}$. This
includes all fully specified $\mathbf{n}$, i.e. with $n^{(abc)} =
(0,0,n)$,
and suffices for the application considered in Section~\ref{sec:ldhat}.
If necessary, $\mathbb{P}_0(\mathbf{n})$ for partially specified $\mathbf{n}$
can be computed by summing over the fully specified configurations consistent with $\mathbf{n}$.

For $|\mathcal{A}|=2$ alleles per locus, 
$\tilde{\Rate}^\timeInd$ is an $O(n^6) \times O(n^6)$ matrix,
so naively computing the matrix multiplication in Theorem~\ref{thm:augmented:moran}
would cost $O(n^{12})$ time.
However, $\tilde{\Rate}^\timeInd$ is sparse, with $O(n^6)$ nonzero entries,
allowing efficient algorithms to compute Theorem~\ref{thm:augmented:moran}
(up to numerical precision) in $O(n^6 \mathcal{T})$ time, where $\mathcal{T}$ is some finite number
of matrix-vector multiplications.
See Appendix~\ref{appendix:action} and \ref{appendix:exact:complexity} for more details.

\subsection{Importance sampling \label{sec:importance:sampling}}

In addition to the approximate \eqref{eq:moran} and exact \eqref{eq:thm1}
formulas for the sampling probability, LDpop includes an importance
sampler for the two-locus ARG $\mathbf{n}_{\leq 0} = \{\mathbf{n}_t\}_{t \leq 0}$.
This provides a method to sample from the posterior distribution of two-locus
ARGs, and also provides an alternative method for computing $\mathbb{P}_0(\mathbf{n})$.
This importance sampler is based on Theorem~\ref{thm:posterior:chain} below,
which characterizes the optimal proposal distribution
for the two-locus coalescent with recombination under variable population size.

Let the proposal distribution $\proposal(\mathbf{n}_{\leq 0})$ be
a probability distribution on $\{\mathbf{n}_{\leq 0} : \mathbf{n}_0 = \mathbf{n}\}$
whose support contains that of $\mathbb{P}(\mathbf{n}_{\leq 0} \mid
\mathbf{n}_0 = \mathbf{n})$. Then we have
\begin{eqnarray*}
  \mathbb{P}_0(\mathbf{n})
&=&
\int_{\mathbf{n}_{\leq 0}: \mathbf{n}_0 = \mathbf{n}}
\frac{d\mathbb{P}(\mathbf{n}_{\leq 0})}{d \proposal(\mathbf{n}_{\leq 0})}
d \proposal(\mathbf{n}_{\leq 0}),
\end{eqnarray*}
and so, if $\mathbf{n}^{(1)}_{\leq 0}, \ldots, \mathbf{n}^{(K)}_{\leq 0} \sim \proposal$ i.i.d., the sum
\begin{eqnarray}
\frac1K \sum_{k=1}^K \frac{d\mathbb{P}(\mathbf{n}^{(k)}_{\leq 0})}{d \proposal(\mathbf{n}^{(k)}_{\leq 0})}
\label{eq:importance:sampling}
\end{eqnarray}
converges
almost surely to $\mathbb{P}_0(\mathbf{n})$ as $K \to \infty$ by the Law of Large Numbers.
Hence, \eqref{eq:importance:sampling} provides a Monte Carlo approximation
to $\mathbb{P}_0(\mathbf{n})$.
The optimal proposal is the posterior distribution $\proposal_{\text{opt}}(\mathbf{n}_{\leq 0}) = \mathbb{P}(\mathbf{n}_{\leq 0} \mid \mathbf{n}_0)$, for then \eqref{eq:importance:sampling} is exactly
\begin{align*}
  \frac1K \sum_{k=1}^K \frac{d\mathbb{P}(\mathbf{n}^{(k)}_{\leq 0})}{d \mathbb{P}(\mathbf{n}^{(k)}_{\leq 0} \mid \mathbf{n}_0)}
&=
  \frac1K \sum_{k=1}^K \frac{d\mathbb{P}(\mathbf{n}^{(k)}_{\leq 0})}{d
    \mathbb{P}(\mathbf{n}^{(k)}_{\leq 0}) / \mathbb{P}(\mathbf{n}_0)}
=
\mathbb{P}(\mathbf{n}_0),
\end{align*}
even for $K=1$.

The following theorem, which we prove in Appendix~\ref{proof:posterior:chain}, characterizes the optimal posterior
distribution $\proposal_{\text{opt}}(\mathbf{n}_{\leq 0})=\mathbb{P}(\mathbf{n}_{\leq 0} \mid \mathbf{n}_0)$ for variable population size:
\begin{theorem} \label{thm:posterior:chain}
The process $\{\mathbf{n}_t\}_{t\leq 0}$ is a backward-in-time Markov chain with inhomogeneous
rates, whose rate matrix at time $t$ is given by
\begin{eqnarray*}
  \isRate^{(t)}_{\mathbf{n}, \mathbf{m}}
&=& 
\begin{cases}
\isQ^{(t)}_{\mathbf{n},\mathbf{m}}
\frac{\mathbb{P}_t(\mathbf{m})}{\mathbb{P}_t(\mathbf{n})},
& \text{if } \mathbf{m} \neq \mathbf{n},
\\
\isQ^{(t)}_{\mathbf{n}, \mathbf{n}} - \frac{d}{dt} \log
\mathbb{P}_t(\mathbf{n}),
& \text{if } \mathbf{m} = \mathbf{n},
\end{cases}
\end{eqnarray*}
where $\isQbold^{(t)}=(\isQ^{(t)}_{\mathbf{n},\mathbf{m}})$ is a square matrix, indexed by configurations
$\mathbf{n}$, with entries given by Table~\ref{tab:Q} and equal to
\begin{eqnarray}
  \isQ^{(t)}_{\mathbf{n}, \mathbf{m}}
&=&
\frac{d}{ds} \Big[\mathbb{P}(n^{(abc)}_{t-s} = m^{(abc)} \mid n^{(abc)}_t =
n^{(abc)}) \,
\notag \\ && \qquad \times
\mathbb{P}(\mathbf{n}_t = \mathbf{n} \mid \mathbf{n}_{t-s} =
\mathbf{m}, n_t^{(abc)} = n^{(abc)}) \Big] \Big|_{s=0}.
\label{eq:Q}
\end{eqnarray}
\end{theorem}

Intuitively, the matrix $\isQbold^{(t)}$ in Table~\ref{tab:Q}
is a linear combination of two rate matrices,
one for propagating $n_t^{(abc)}$ an infinitesimal distance
backwards in time, and another for propagating $\mathbf{n}_t$ an infinitesimal distance forward in time.
This is because $n_t^{(abc)}$ is generated by sampling coalescent and recombination events backwards in time, and then $\mathbf{n}_t$ is generated by dropping mutations on the ARG and propagating the allele values forward in time.

Theorem~\ref{thm:posterior:chain} generalizes previous results for
the optimal proposal distribution in the constant size case \citep{stephens2000inference, fearnhead2001estimating}.
In that case, the conditional probability of the parent
$\mathbf{m}$ of $\mathbf{n}$ is
$\isQ_{\mathbf{n},\mathbf{m}}
\frac{\mathbb{P}(\mathbf{m})}{\mathbb{P}(\mathbf{n})}$.
Note the constant size case is time-homogeneous, so the dependence on $t$ is dropped,
and the waiting times between events in the ARG are not sampled
(i.e., only the embedded jump chain of $\mathbf{n}_{\leq 0}$ is sampled).

\begin{table}[t]
  \caption{Nonzero entries of the $\isQbold^{(t)}$ matrix of
    Theorem~\ref{thm:posterior:chain}, for $t \in (t_{\timeInd},
    t_{\timeInd+1}]$.}
\label{tab:Q}
  \centering
  \begin{tabular}{cc}
\hline
 $\mathbf{m}$ &
 $\isQ^{(t)}_{\mathbf{n},\mathbf{m}}$
\\ \hline 
$\mathbf{n} - \mathbf{e}_{i*} + \mathbf{e}_{j*}$
&
$\frac{\theta}2 P_{ji} (n_{j*}+1)$
\\ 
$\mathbf{n} - \mathbf{e}_{*i} + \mathbf{e}_{*j}$
&
$\frac{\theta}2 P_{ji} (n_{*j}+1)$
\\ 
$\mathbf{n} - \mathbf{e}_{ij} + \mathbf{e}_{kl}$
&
$\frac{\theta}2 (\delta_{ik} P_{lj} + \delta_{jl} P_{ki}) (n_{kl}+1)$
\\ 
$\mathbf{n} - \mathbf{e}_{ij} + \mathbf{e}_{i*} + \mathbf{e}_{*j}$
&
$\frac{\rho}2 n^{(c)} (n_{i*} + 1) (n_{*j} + 1)$
\\ 
$\mathbf{n} - \mathbf{e}_{ij}$
& 
$\frac1{\eta_\timeInd} {n^{(c)} \choose 2} (n_{ij} - 1)$
\\ 
$\mathbf{n} - \mathbf{e}_{i*}$
&
$\frac1{\eta_\timeInd} \big[ {n^{(a)} \choose 2}(n_{i*}-1) + n^{(a)} n^{(c)} \sum_j n_{ij}\big]$
\\ 
$\mathbf{n} - \mathbf{e}_{*i}$
&
$\frac1{\eta_\timeInd} \big[{n^{(b)} \choose 2}(n_{*i}-1) + n^{(b)} n^{(c)} \sum_j n_{ji}\big]$
\\ 
$\mathbf{n}$
& $-\frac1{\eta_\timeInd} {n \choose 2} - \frac{\rho}2 n^{(c)} - \frac{\theta}2 \sum_i (1-P_{ii}) [n_{i*} + n_{*i} + \sum_j (n_{ij} + n_{ji})] $
\\ \hline
  \end{tabular}
\end{table}

We construct our proposal distribution
$\hat{\proposal}(\mathbf{n}_{\leq 0})$
by approximating the optimal proposal distribution
$\proposal_{\text{opt}}(\mathbf{n}_{\leq 0}) =
\mathbb{P}(\mathbf{n}_{\leq 0} \mid \mathbf{n}_0)$.
This requires approximating the rate 
$\isRate^{(t)}_{\mathbf{n}, \mathbf{m}} = \isQ^{(t)}_{\mathbf{n},\mathbf{m}} \frac{\mathbb{P}_t(\mathbf{m})}{\mathbb{P}_t(\mathbf{n})}$.
We use the approximation $\hat{\isRate}^{(t)}_{\mathbf{n},\mathbf{m}} = \isQ^{(t)}_{\mathbf{n},\mathbf{m}} \frac{\mathbb{P}^{(N)}_t(\mathbf{m})}{\mathbb{P}^{(N)}_t(\mathbf{n})}$, with $\mathbb{P}^{(N)}_t(\mathbf{n})$ from the approximate likelihood formula
\eqref{eq:moran}.
To save computation, we only compute $\mathbb{P}^{(N)}_t(\mathbf{n})$ along a grid of time points,
and then linearly interpolate $\hat{\isRate}^{(t)}_{\mathbf{n},\mathbf{m}}$ between the points.
See Appendix~\ref{appendix:proposal} for more details on our proposal distribution $\hat{\proposal}$.

As detailed in Section~\ref{sec:is:accuracy:runtime}, $\hat{\proposal}$ is a highly efficient proposal distribution, typically yielding
effective sample sizes (ESS) between 80 to 100\% per sample for the demography and $\rho$ values we considered.

\section{Application \label{sec:empirical}}

Previous simulation studies \citep{mcvean2002coalescent,chan2012genome,smith2005comparison} have shown that if the demographic model is misspecified, composite-likelihood methods (which so far have assumed a constant population size) can produce recombination rate estimates that are biased.
Many populations, including that of humans and \emph{D. melanogaster}, have undergone bottlenecks and  expansions in the recent past \citep{gutenkunst2009inferring, choudhary1987historical}, and it has been argued \citep{johnston2012population} that such demographies can severely affect recombination rate estimation, and can cause the appearance of spurious recombination hotspots.

In this section, we apply our software LDpop to show 
that accounting for demography improves fine-scale recombination rate estimation.
We first examine how a population bottleneck followed by rapid growth affects the correlation between partially linked sites.
We then study 
composite likelihood estimation of recombination maps
under a population bottleneck.  
We find that accounting for demography with either the exact (Theorem~\ref{thm:augmented:moran})
or approximate \eqref{eq:moran} likelihood formula substantially
improves accuracy.
Furthermore, this improvement is robust to minor misspecification of the demography
due to not knowing the true demography in practice.

Throughout this section, we use an example demography with $\pieces=3$ epochs, consisting of a sharp population bottleneck followed by a rapid expansion.
Specifically, the population size history $\eta(t)$, in coalescent-scaled units, is given by
\begin{equation}
\eta(t) = \begin{cases} 
      100, &  -0.5 < t \leq 0, \\
      0.1, & -0.58 < t \leq -0.5, \\
      1, & t \leq -0.58.
   \end{cases}
\label{eq:exampleDemo}
\end{equation}
Under this model and $n=2$, the expected time of common ancestor is $\mathbb{E}[T_{\text{MRCA}}] \approx 1$.
We thus compare this demography against a constant size demography
with coalescent-scaled size of $\eta \equiv 1$, as this is the population size that would be estimated
using the pairwise heterozygosity \citep{tajima1983evolutionary}.
We use a coalescent-scaled mutation rate of $\frac{\theta}2=0.004$ per base, which is roughly
the mutation rate of \emph{D. melanogaster} \citep{chan2012genome}.

While the size history $\eta(t)$ of \eqref{eq:exampleDemo} is fairly simple, with
only $\mathcal{D}=3$ epochs, we stress that LDpop can in fact handle much more complex size histories.
For example, in Section~\ref{sec:runtime}, we show that LDpop
can easily handle a demography with $\mathcal{D}=64$, with little additional cost in runtime.

\begin{figure}[t]
\def\svgwidth{.5\textwidth}
\centerline{\includegraphics[width=0.5\textwidth]{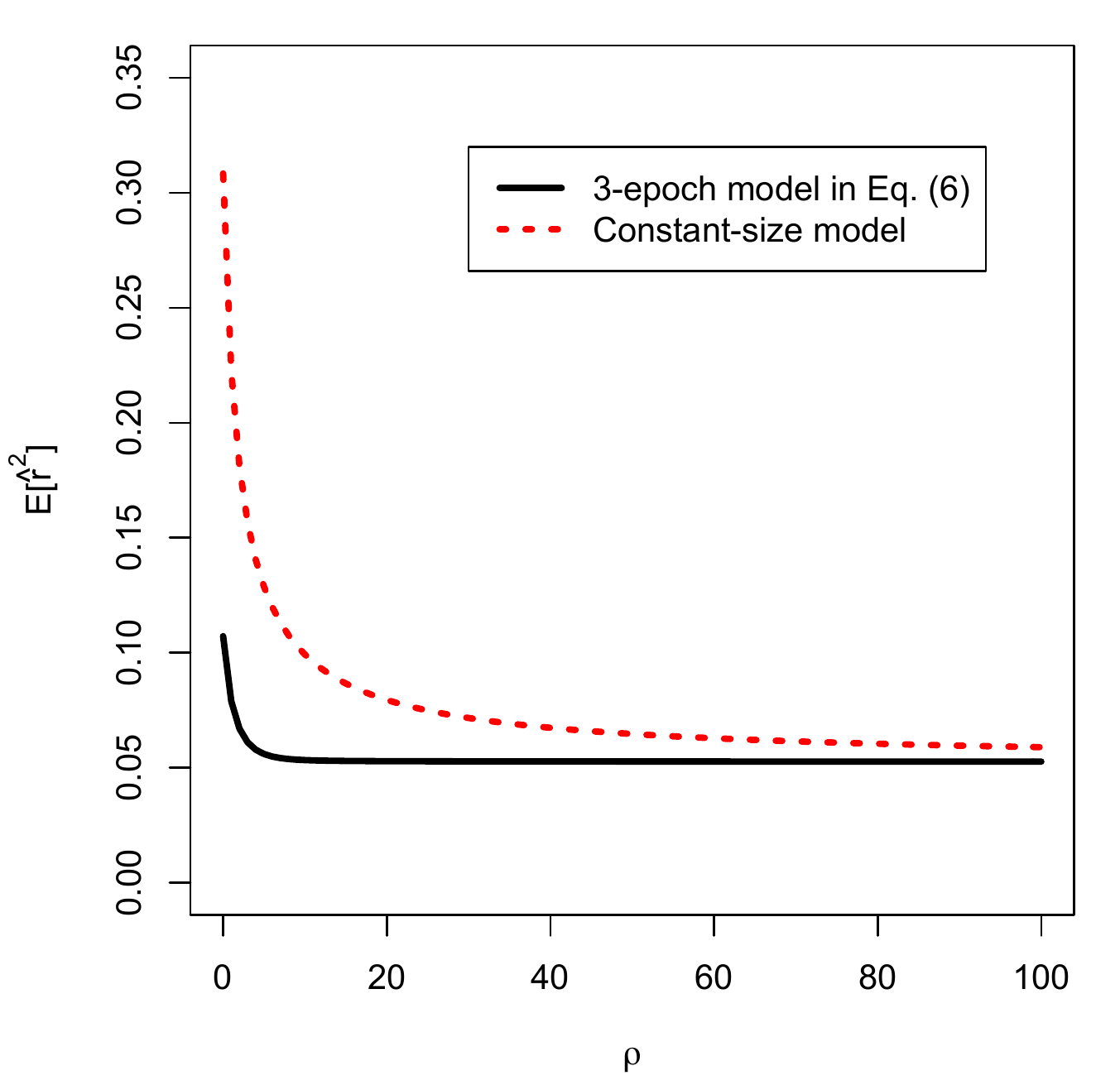}}
\caption{Expected linkage disequilibrium $\mathbb{E}[\hat{r}^2]$ for the 3-epoch model in \eqref{eq:exampleDemo} and the constant size model with $\eta \equiv 1$, as a function of recombination rate $\rho$ for a sample size of $n=20$.
Under the 3-epoch model, even nearby sites are expected to be quite uncorrelated.}
\label{fig:r2}
\end{figure}

\subsection{Linkage disequilibrium and two-locus likelihoods\label{sec:ld}}

One statistic of linkage disequilibrium is
\begin{align*}
  \hat{r}^2 &= \frac{\big[\hat{x}_{11}-\hat{x}^{(a)}_{1} \hat{x}^{(b)}_{1}\big]^2}{\hat{x}^{(a)}_{0} \hat{x}^{(a)}_{1} \hat{x}^{(b)}_{0} \hat{x}^{(b)}_{1}},
\end{align*}
where $\hat{x}_{ij} = \frac{n_{ij}}n$ is the fraction of the sample with haplotype $ij$, with $\hat{x}^{(a)}_i = \sum_j \hat{x}_{ij}$ and  $\hat{x}_j^{(b)} = \sum_i \hat{x}^{(b)}_{ij}$.
In words, $\hat{r}^2$ is the sample square-correlation of a random allele at locus $a$ with a random allele at locus $b$.
We let $r^2 = \lim_{n \to \infty} \hat{r}^2$ denote the population square-correlation.
There has been considerable theoretical interest in understanding moments of $r^2$, $\hat{r}^2$, and related statistics
\citep{ohta1969linkage, maruyama1982stochastic, hudson1985sampling, mcvean2002genealogical, song2007analytic}.
Additionally, $n \hat{r}^2$ approximately follows a $\chi^2_1$ distribution when $r^2 = 0$, 
which provides a test for the statistical significance of linkage disequilibrium \citep[p.~113]{weir1996genetic}.

Using LDpop, we can compute the distribution of $\hat{r}^2$ for piecewise constant models.
In Figure~\ref{fig:r2}, we show $\mathbb{E}[\hat{r}^2]$ for a sample size $n = 20$ under the 3-epoch model \eqref{eq:exampleDemo} and the constant population size model.
Under the 3-epoch demography,
$\mathbb{E}[\hat{r}^2]$ is much lower for small $\rho$ and decays more rapidly as $\rho \to \infty$. 
In other words, the constant model requires higher $\rho$ to break down LD to the same level, which suggests
that incorrectly assuming a constant demography will lead to upward-biased estimates of the recombination rate (as pointed out by an anonymous reviewer).
We confirm this in Section~\ref{sec:ldhat}.

\subsection{Fine-scale recombination rate estimation \label{sec:ldhat}}

For a sample of $n$ haplotypes observed at $L$ SNPs,
let $\mathbf{n}[a,b]$ be the two-locus sample observed at SNPs $a,b \in \{1,\ldots,L\}$,
and let $\rho[a,b]$ be the recombination rate between SNPs $a$ and $b$.
The programs LDhat \citep{mcvean2002coalescent, mcvean2004fine, auton2007recombination}, LDhot \citep{myers2005fine,auton2014identifying}, and LDhelmet \citep{chan2012genome} infer hotspots and recombination maps $\bfmath\rho$ using the following composite likelihood due to \citet{hudson2001two}:
\begin{align}
\prod_{a,b: \ 0< b-a < W} \mathbb{P}(\mathbf{n}[a,b] ; \rho[a,b])
\label{eq:composite:likelihood:noroot}
\end{align}
where $W$ denotes some window size in which to consider pairs of sites (a finite window size $W$ removes the computational burden and statistical noise from distant, uninformative sites \citep{fearnhead2003consistency, smith2005comparison}).

We used LDpop to generate four likelihood tables, which we then used with LDhat and LDhelmet to
estimate recombination maps for simulated data. The four tables, denoted by $\Lconst$, $\Lexact$, $\Lapprox$,
and $\Lmiss$, are defined as follows:
\begin{enumerate}
\item (``Constant'') $\Lconst$ denotes a likelihood table that assumes a constant population size of $\eta \equiv 1$.
\item (``Exact'') $\Lexact$ denotes a likelihood table that assumes the correct 3-epoch population size
history $\eta$ defined in \eqref{eq:exampleDemo}.
\item (``Approximate'') $\Lapprox$ denotes a likelihood table
with the correct size history $\eta$ in \eqref{eq:exampleDemo}, but using the (much faster) approximate likelihood formula \eqref{eq:moran} with $N=n$ Moran particles.
\item (``Misspecified'') $\Lmiss$ denotes a likelihood table that assumes a misspecified demography $\hat{\eta}$, defined by
\begin{equation}
\hat{\eta}(t) = \begin{cases} 
      90.5, &  -0.534 < t \leq 0, \\
      0.167, & -0.66 < t \leq -0.534, \\
      1.0, & t \leq -0.66,
   \end{cases}
\label{eq:missDemo}
\end{equation}
which was estimated from simulated data (Appendix~\ref{appendix:misspecified})
\end{enumerate}

Overall, we found that using the constant table $\Lconst$ leads to very noisy and biased estimates of $\bfmath{\rho}$ (as might be expected from Figure~\ref{fig:r2}).
The other tables $\Lexact$, $\Lapprox$, and $\Lmiss$ all lead to much more accurate estimates.
Using $\Lexact$ (the exact likelihood table with the true size history) produces slightly more accurate estimates of $\bfmath{\rho}$ than using $\Lapprox$ or $\Lmiss$.
However, the three non-constant tables $\Lexact$, $\Lapprox$, and $\Lmiss$ all produce very similar results that are hard to distinguish from one another.

\begin{figure}[p]
\centering
\begin{subfigure}[t]{.44\textwidth}
\includegraphics[width=\textwidth]{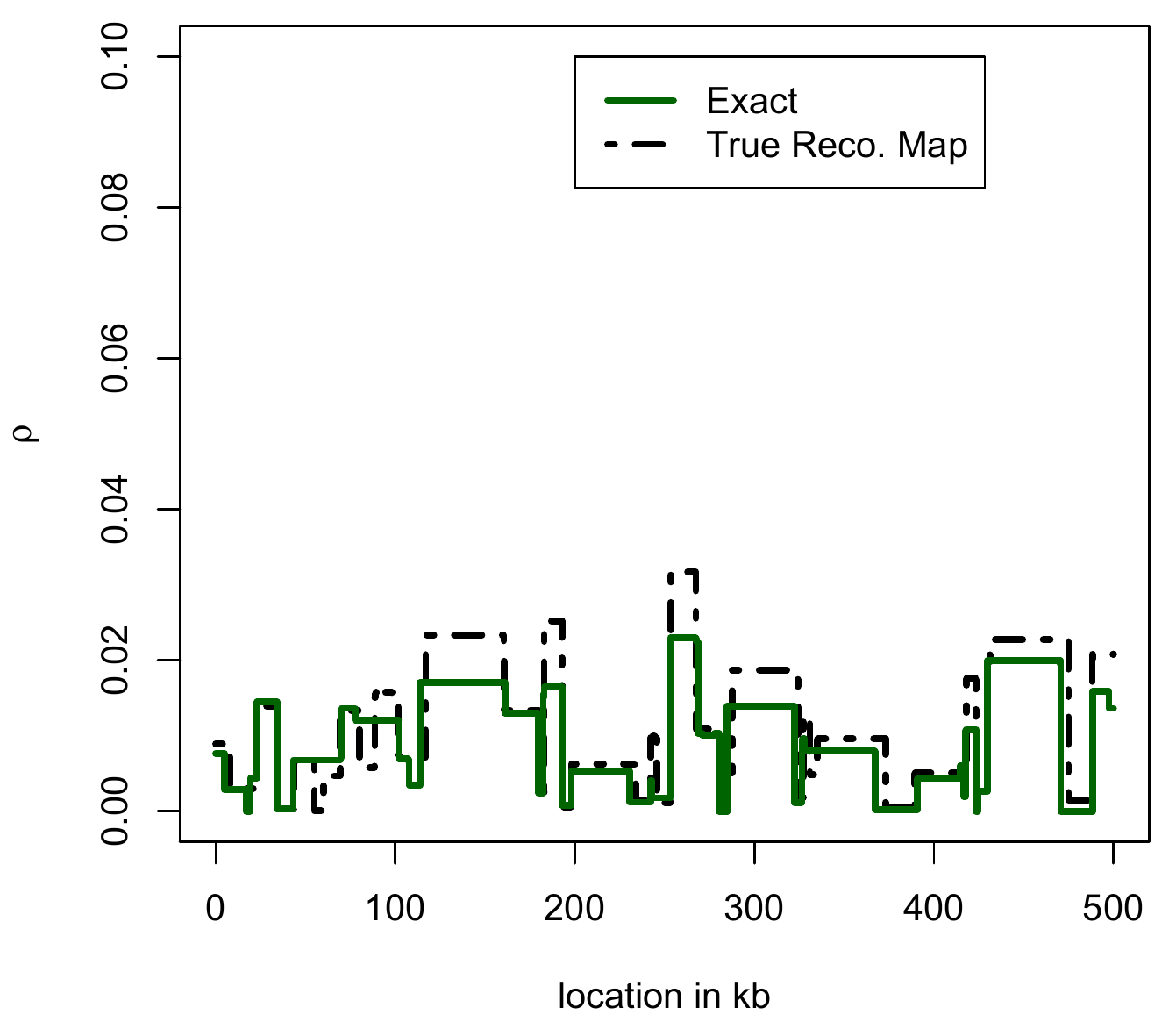}
\caption{}
\end{subfigure}
\begin{subfigure}[t]{.44\textwidth}
\includegraphics[width=\textwidth]{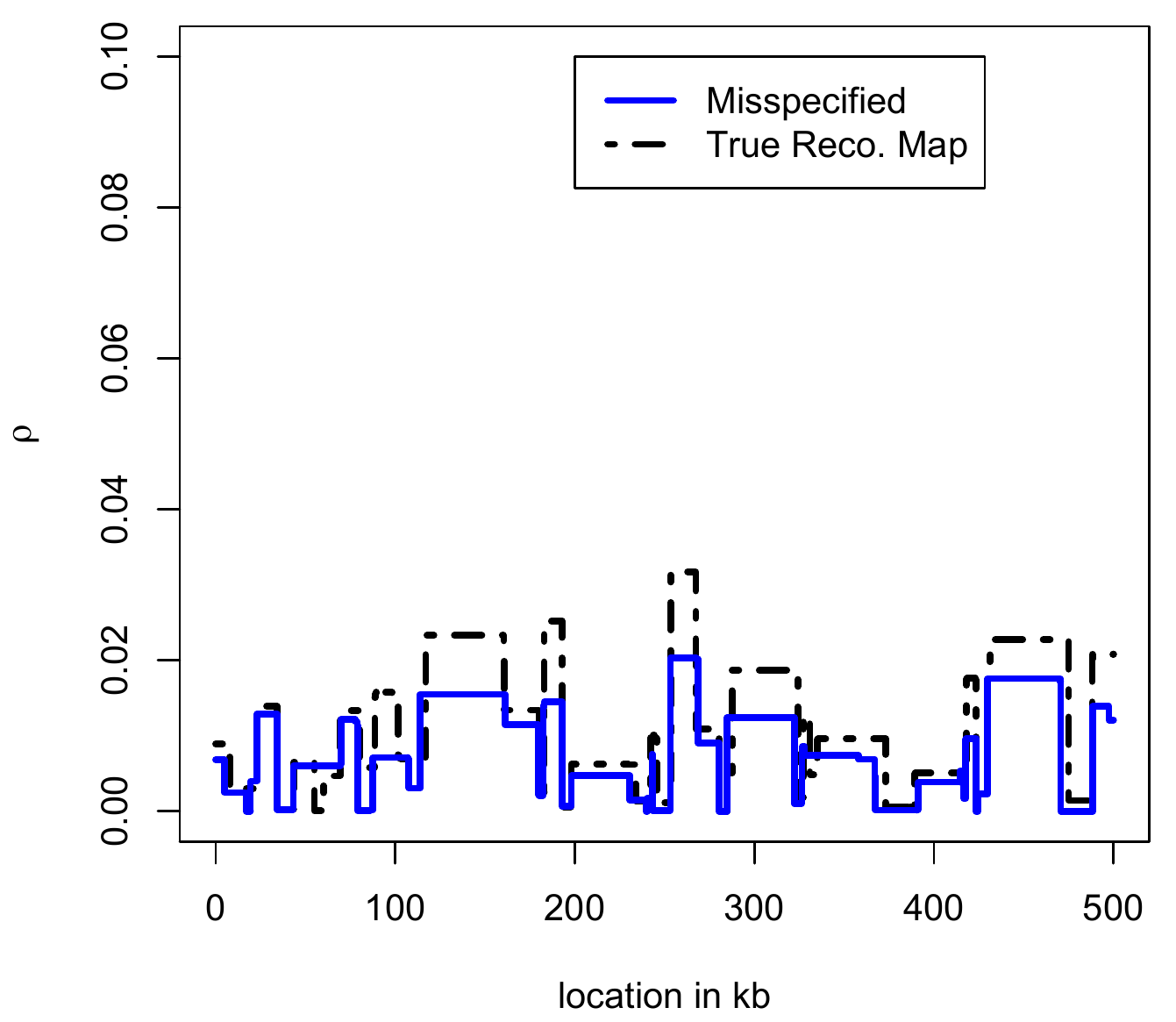}
\caption{}
\end{subfigure}
\begin{subfigure}[t]{.44\textwidth}
\includegraphics[width=\textwidth]{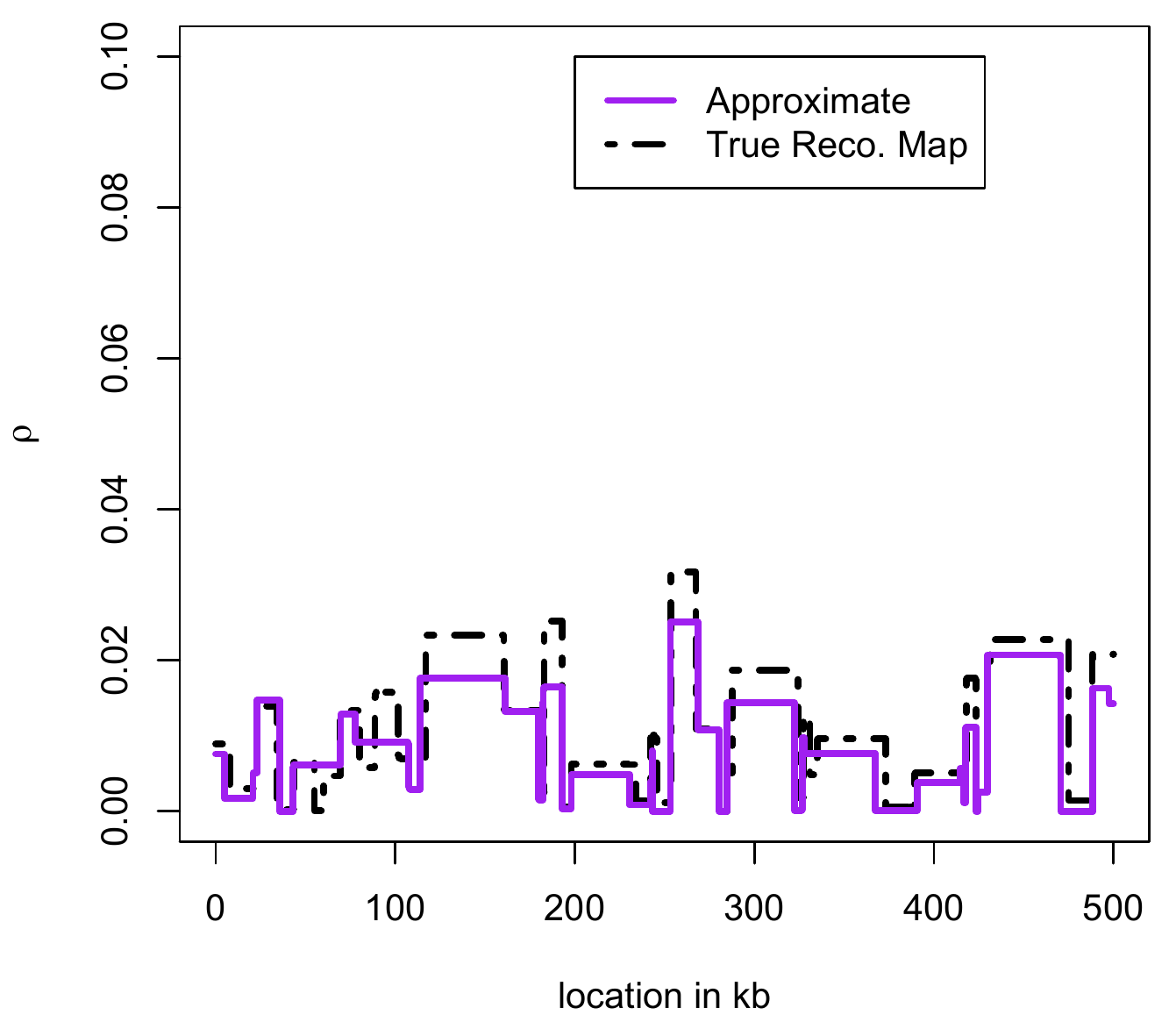}
\caption{}
\end{subfigure}
\begin{subfigure}[t]{.44\textwidth}
\includegraphics[width=\textwidth]{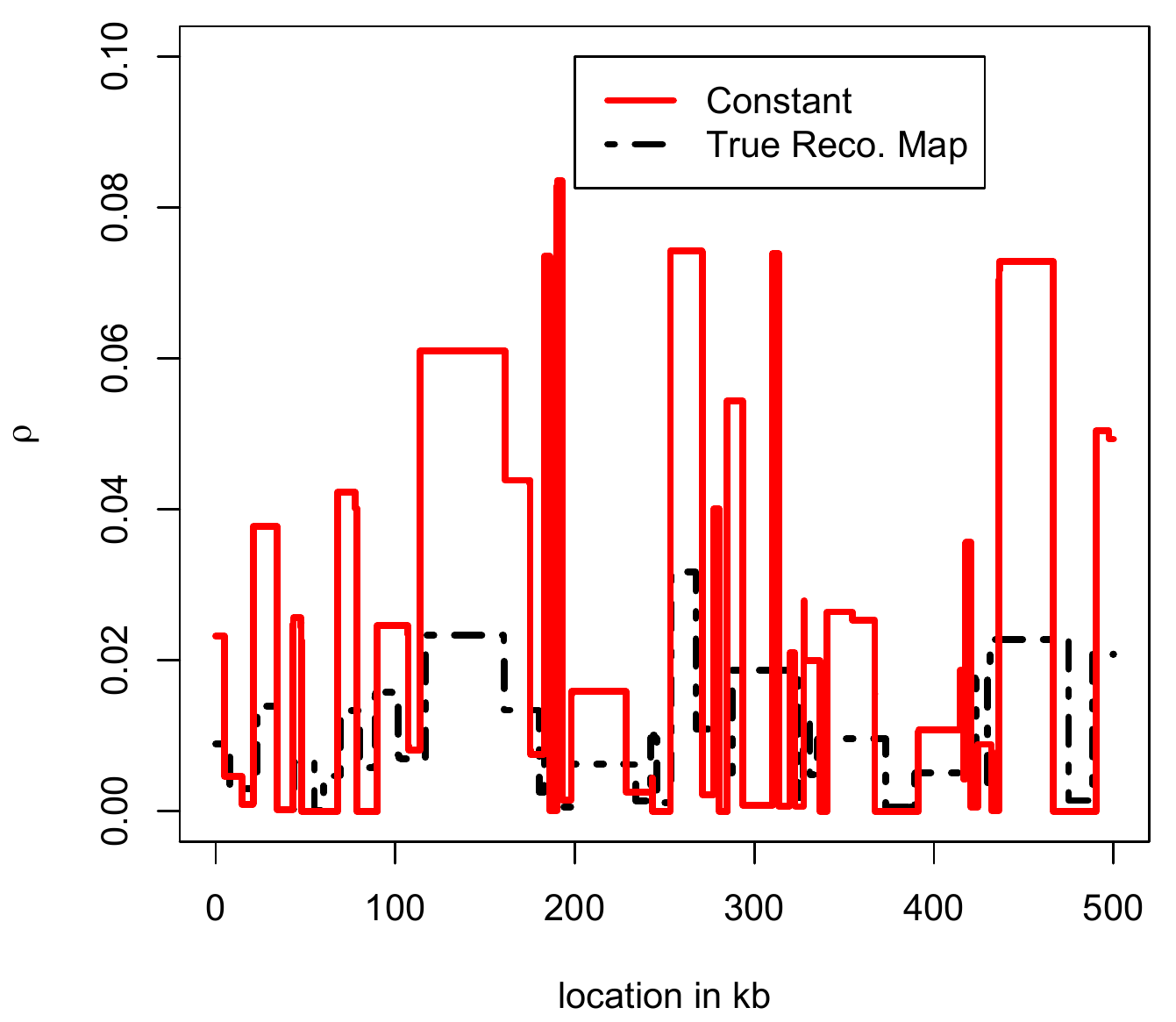}
\caption{}
\end{subfigure}
\caption{Comparison of recombination maps inferred using different lookup tables.
We simulated $n=20$ haplotypes under the 3-epoch model \eqref{eq:exampleDemo} and using the recombination map shown in black dashed line.
(a) Inferred map $\bfmath{\hat{\rho}}_{\Lexact}$ obtained using the exact likelihoods for the true demography.
(b) Inferred map $\bfmath{\hat{\rho}}_{\Lmiss}$ obtained using the empirically estimated demography in \eqref{eq:missDemo}.
(c) Inferred map $\bfmath{\hat{\rho}}_{\Lapprox}$ obtained using an approximate lookup table for the true demography.
(d) Inferred map $\bfmath{\hat{\rho}}_{\Lconst}$ obtained assuming a constant population size of $\eta\equiv 1$.
Note that $\bfmath{\hat{\rho}}_{\Lconst}$ is much noisier than the other estimates, while using an inferred demography or an approximate lookup table results in only a very mild reduction in accuracy compared to using the true sampling probabilities. 
 These $\bfmath{\hat{\rho}}$ were produced with LDhelmet; using LDhat led to very similar results.}
\label{fig:representative:rho}
\end{figure}

\begin{figure}[p]
\centering
\begin{subfigure}[t]{.55\textwidth}
  \includegraphics[width=\textwidth]{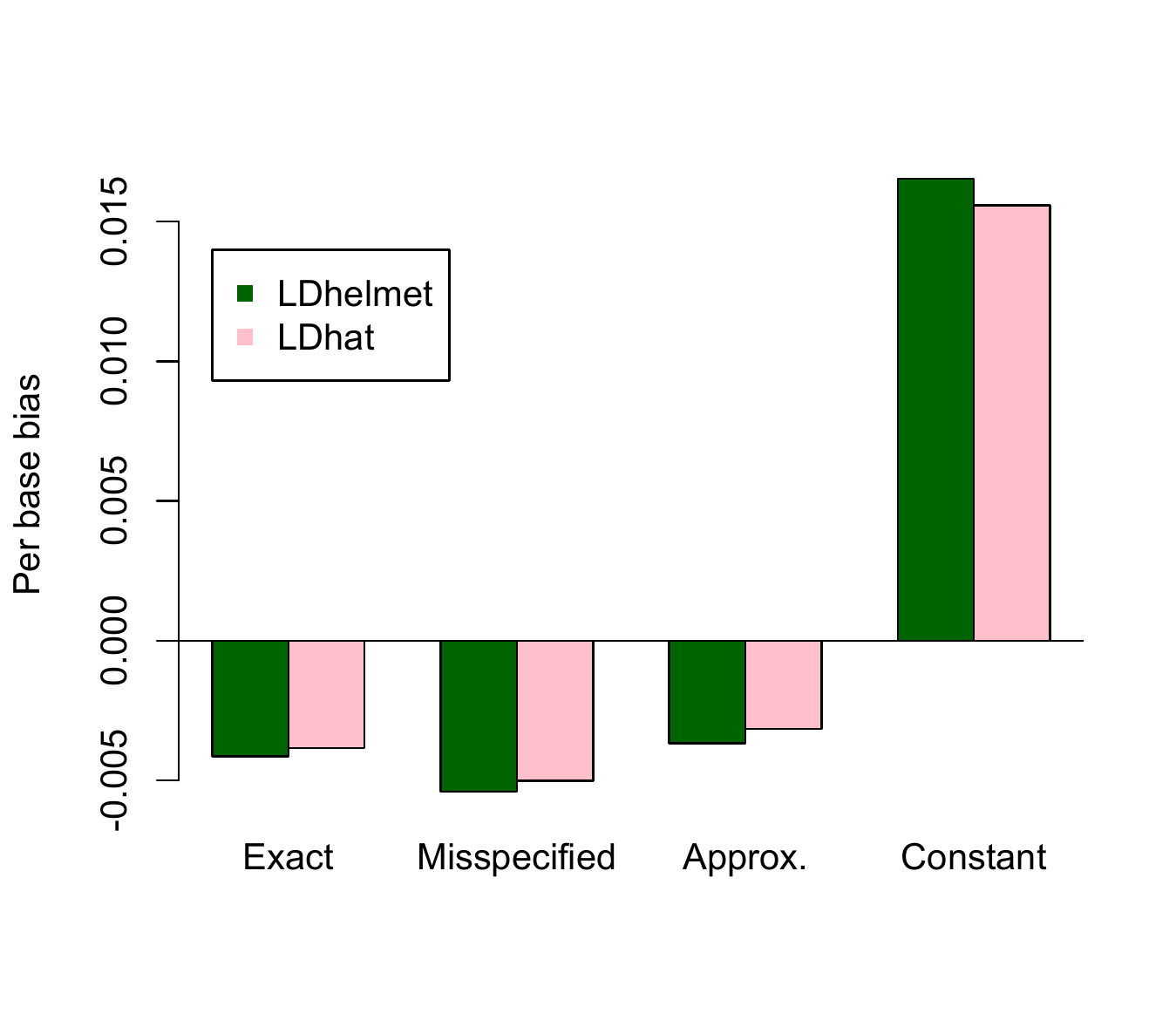}
\caption{}
\end{subfigure} 
\hspace{10mm}
\begin{subfigure}[t]{.55\textwidth}
\includegraphics[width=\textwidth]{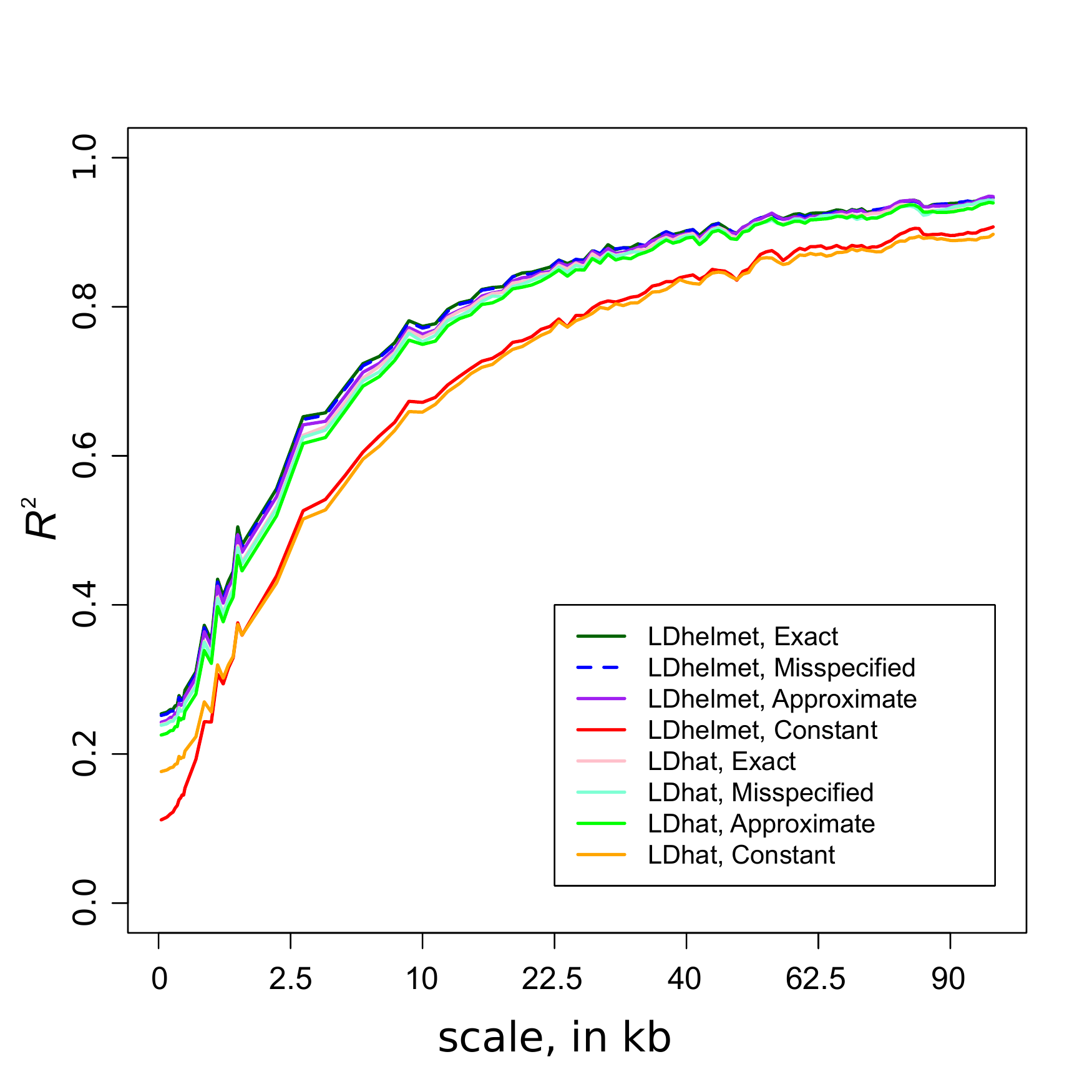}
\caption{}  
\end{subfigure}
\caption{Accuracy of the estimated maps $\bfmath{\hat{\rho}}_{\Lexact}, \bfmath{\hat{\rho}}_{\Lmiss}, \bfmath{\hat{\rho}}_{\Lapprox},\bfmath{\hat{\rho}}_{\Lconst}$ over 110 simulations similar to Figure~\ref{fig:representative:rho}. 
The estimate $\bfmath{\hat{\rho}}_{\Lconst}$ obtained assuming constant demography is substantially more biased and noisier than the other estimates.
(a) The average per-base bias $\bfmath{\hat{\rho}}-\bfmath{\rho}$.
(b) The square Pearson correlation coefficient $R^2$ over different scales.  This $R^2$ is distinct from the $r^2$ statistic measuring linkage disequilibrium (Section~\ref{sec:ld}). To compute $R^2$ for scale $s$, the middle 500~kb region of each 1~Mb simulation was divided into non-overlapping windows of size $s$ and we compared the average of $\bfmath{\hat{\rho}}$ to the average of $\bfmath{\rho}$  in each window. The $x$-axis is stretched by $x \mapsto \sqrt{x}$.
}
\label{fig:r2:all}
\end{figure}

Figures~\ref{fig:representative:rho} and \ref{fig:r2:all} show the accuracy
of estimated recombination maps $\bfmath{\hat{\rho}}$ on simulated data.
We simulated $n=20$ sequences under the 3-epoch demography defined in \eqref{eq:exampleDemo},
with the true maps $\bfmath{\rho}$ taken from previous estimates
for the X chromosome of \emph{D. melanogaster} \citep{chan2012genome}.
In all, there were 110 independent datasets, with estimated maps $\bfmath{\hat{\rho}}$
of length 500~kb. See Appendix~\ref{appendix:details} for further details.

In Figure~\ref{fig:representative:rho}, we plot $\bfmath{\rho}$ and $\bfmath{\hat{\rho}}$ for a particular 500~kb region.
Qualitatively, the constant size estimate $\bfmath{\hat{\rho}}_{\Lconst}$ is less accurate and has wilder fluctuations.
Figure~\ref{fig:r2:all} shows that over all 110 replicates, the constant size estimate $\bfmath{\hat{\rho}}_{\Lconst}$ has high bias and low correlation with the truth,
compared to the estimates $\bfmath{\hat{\rho}}_{\Lexact}$, $\bfmath{\hat{\rho}}_{\Lmiss}$, and $\bfmath{\hat{\rho}}_{\Lapprox}$ which account for variable demography.
Following \citet{wegmann2011recombination}, we plot the correlation of 
$\bfmath{\hat{\rho}}$ with $\bfmath{\rho}$ at multiple scales; at all scales, the constant-demography estimate $\bfmath{\hat{\rho}}_{\Lconst}$ is considerably worse than the other estimates.
In general, using an inferred demography or an approximate lookup table results in only a very mild reduction in accuracy compared to using the true sampling probabilities.

\begin{figure}[p]
    \centering
    \begin{subfigure}[t]{0.44\textwidth}
\includegraphics[width=\textwidth]{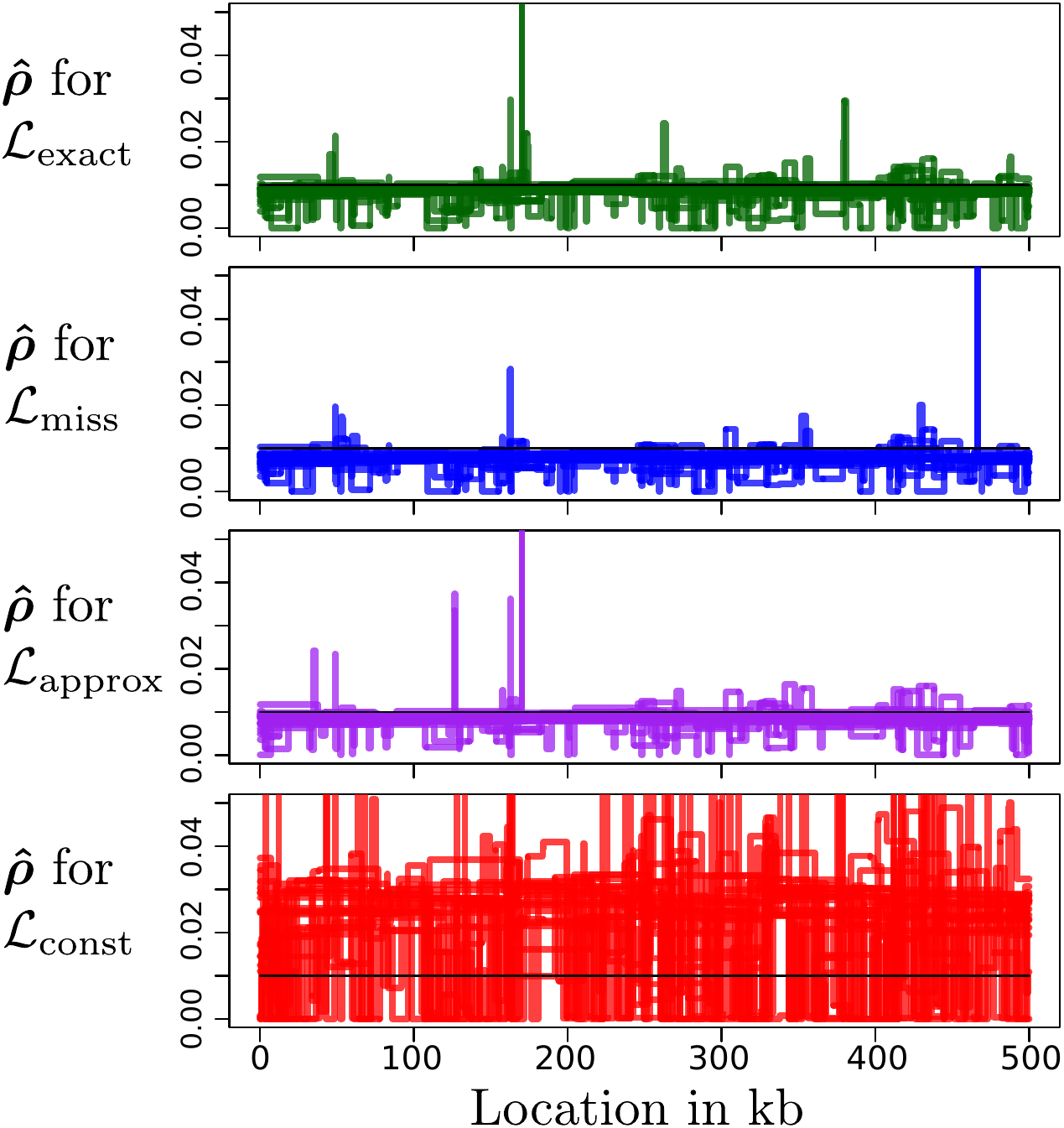}
\caption{}
    \end{subfigure}
    \qquad \qquad 
    \begin{subfigure}[t]{0.44\textwidth}
\includegraphics[width=\textwidth]{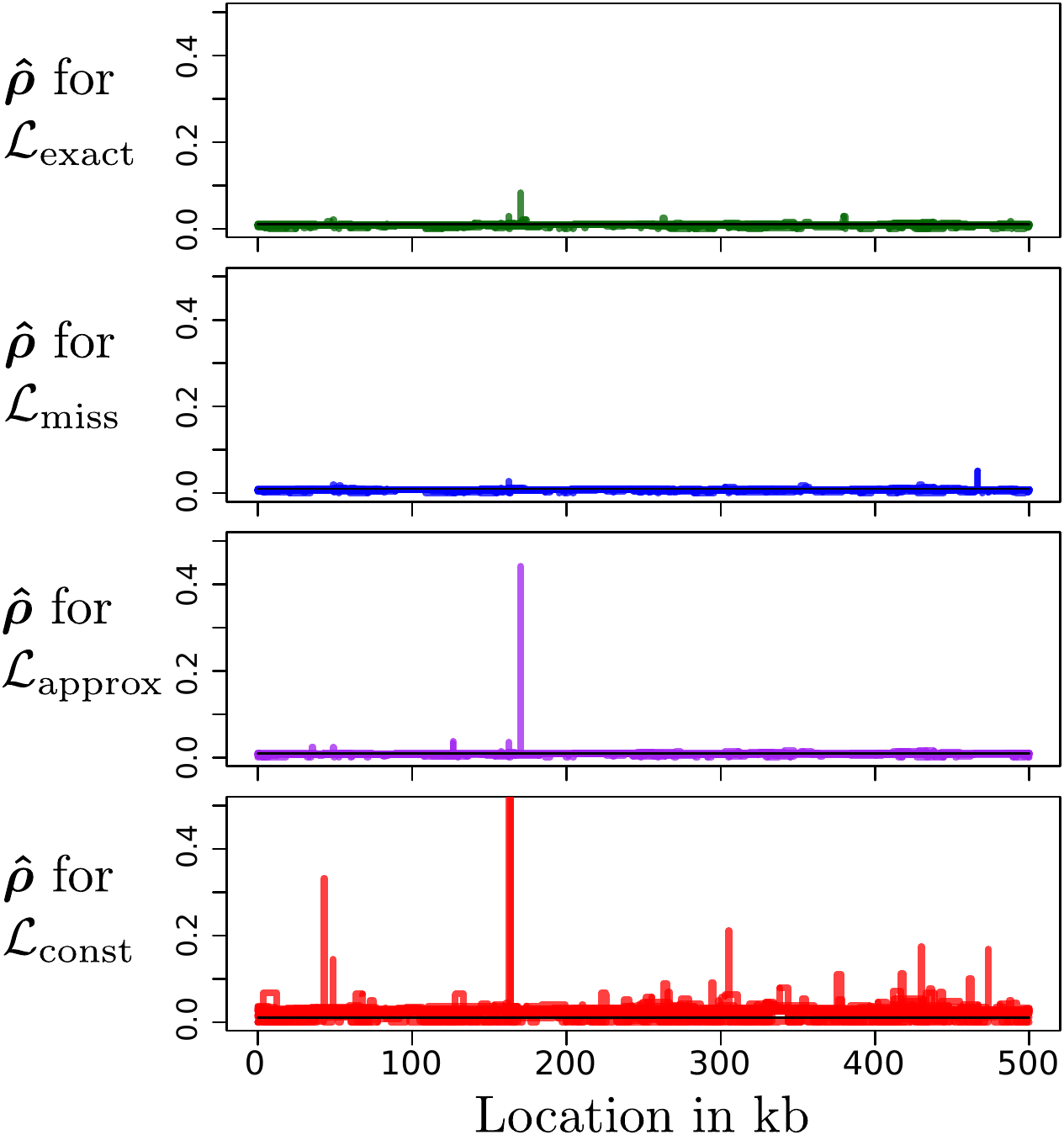}
\caption{}
    \end{subfigure}
\caption{Inferred recombination maps $\bfmath{\hat{\rho}}$ produced by LDhelmet using four different look up tables ($\Lconst, \Lexact, \Lapprox,\Lmiss$), when the true recombination rate is constant at $\rho = 0.01$ per bp.
Results for 20 simulated datasets are shown.  Each simulation was done under the 3-epoch demography defined in \eqref{eq:exampleDemo}.
Results for LDhat were very similar (not shown). 
(a) Inferred recombination maps $\bfmath{\hat{\rho}}$.
(b) The same plot but zoomed out.  For the figure corresponding to $\Lconst$, the highest peak reaches $\rho = 2.5$ (not displayed), which is 250 times the true value.
}
\label{fig:constant:rho}
\end{figure}

We also considered a constant (flat) map $\bfmath{\rho}$;
the estimated $\bfmath{\hat{\rho}}$ are shown in Figure~\ref{fig:constant:rho}.
Consistent with \citet{johnston2012population}, we find that the constant-demography
estimate $\bfmath{\hat{\rho}}_{\Lconst}$ can have extreme peaks, and is generally
very noisy.
On the other hand, the estimates $\bfmath{\hat{\rho}}_{\Lexact},\bfmath{\hat{\rho}}_{\Lmiss},\bfmath{\hat{\rho}}_{\Lapprox}$ that account for demography
have less noise and fewer large deviations.

\section{Runtime and Accuracy of Likelihoods} \label{sec:runtime}

\subsection{Runtime of the exact and approximate likelihood formulas\label{sec:runtime:exact:approx}}

Both the approximate \eqref{eq:moran} and exact likelihood formulas (Theorem~\ref{thm:augmented:moran})
require computing products $ \mathbf{v} e^{\bfA}$ for some $k \times k$ matrix $\bfA$ and $1 \times k$ row vector $\mathbf{v}$.
Naively, this kind of vector-matrix multiplication costs $O(k^2)$.
However, in our case $\bfA$ is sparse, with $O(k)$ nonzero entries,
allowing us to compute $\mathbf{v} e^{\bfA}$ up to numerical precision
in $O(k \mathcal{T})$ time, where $\mathcal{T}$ is some finite number of sparse
matrix-vector products depending on $\bfA$ \citep{al-mohy2011computing}.
In particular, $k=O(n^3)$ for the approximate formula, whereas $k=O(n^6)$ for the exact
formula.
Thus, computing the likelihood table costs $O(n^3 \mathcal{T})$ and $O(n^6 \mathcal{T})$
for the approximate \eqref{eq:moran} and exact (Theorem~\ref{thm:augmented:moran}) formulas,
respectively.
See Appendix~\ref{appendix:complexity} for a more detailed analysis of the
computational complexity, and a description of the algorithm for computing 
$\mathbf{v} e^{\bfA}$.
Note that $\mathcal{T}$ depends nontrivially on the sample size $n$,
as well as the parameters $\rho$, $\theta$, $\eta_d$, and $t_d$.

\begin{figure}[p]
  \centering
    \begin{subfigure}[t]{0.5\textwidth}
  \includegraphics[width=\textwidth]{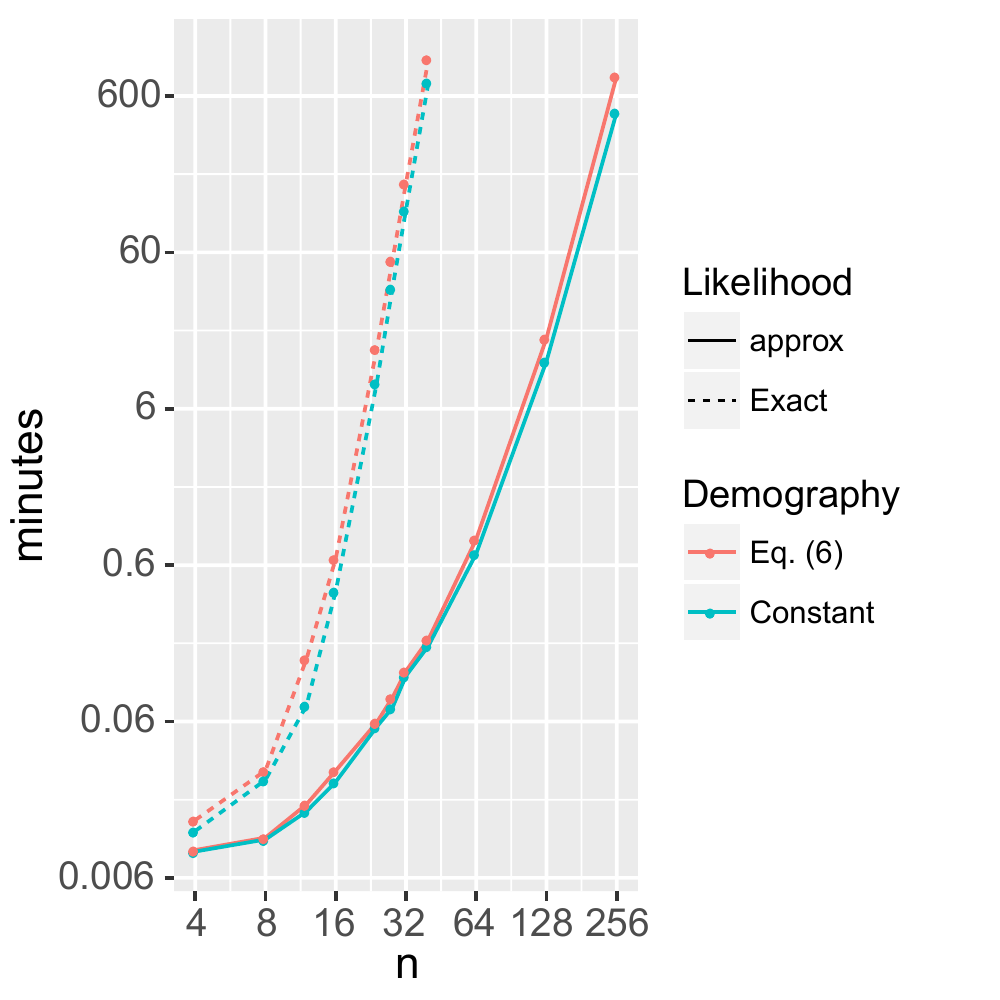}
        \caption{}
    \end{subfigure}
    \qquad \qquad 
    \begin{subfigure}[t]{0.6\textwidth}
  \includegraphics[width=\textwidth]{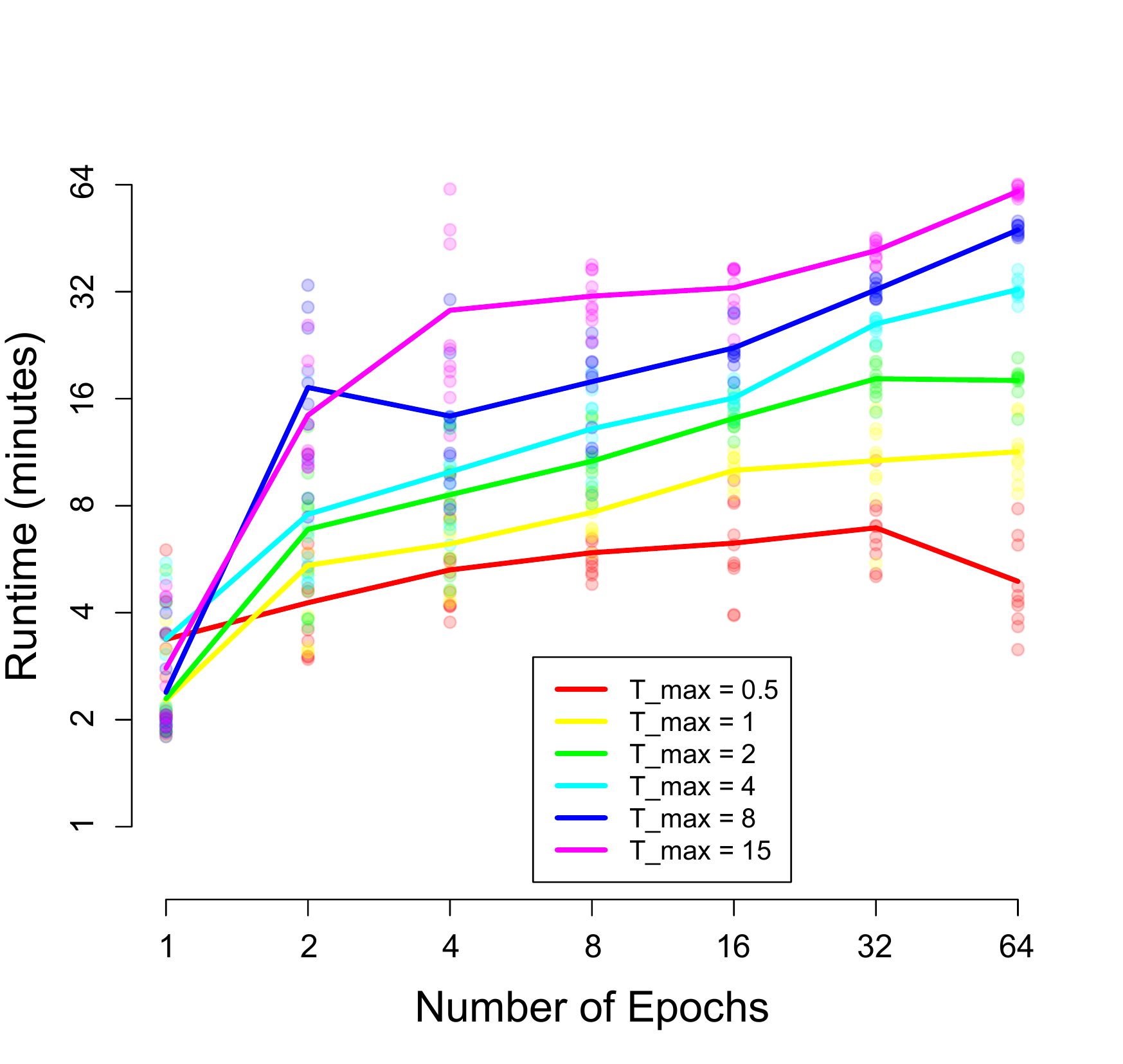}
        \caption{}
    \end{subfigure}
  \caption{Runtime of LDpop to compute a likelihood table with $\rho \in \{0,1,\ldots,100\}$. Experiments were performed using 24 cores on a computer with 256~GB of RAM. (a) Runtime as a function of sample size, for both the approximate and exact formulas, and for two demographies (a constant size history and the 3-epoch model \eqref{eq:exampleDemo}).
  (b) Runtime as a function of the number of epochs $\mathcal{D}$ on $[-T_{\max},0]$, for the exact formula with $n=20$, with 10 repetitions with random population sizes. Notice that runtime depends more on $T_{\max}$ than $\mathcal{D}$, and grows sublinearly with $\mathcal{D}$ (a $2$-fold increase of $\mathcal{D}$ yields less than a $2$-fold increase in runtime).
  }
  \label{fig:runtime}
\end{figure}

We present running times for LDpop in Figure~\ref{fig:runtime}.
We used LDpop to generate likelihood tables with $\rho \in \{0,1,\ldots,100\}$,
 using 24 cores on a computer with 256~GB of RAM.
On the 3-epoch demography \eqref{eq:exampleDemo},
we ran the exact formula up to $n=40$ (17 hours), and the approximate formula
up to $n=256$ (13 hours).
The constant demography takes nearly the same amount of time as the 3-epoch
demography, which agrees with our general experience that computing the initial stationary distribution
$\stationaryLambdaTilde{-\pieces}$ is more expensive than multiplying the matrix exponentials
$e^{\tilde{\Rate}^\timeInd t}$. Using faster algorithms to compute $\stationaryLambdaTilde{-\pieces}$
should lead to substantial improvements in runtime.

Figure~\ref{fig:runtime} also examines how LDpop scales with the number of epochs in the demography.
We split $[-T_{\max},0]$ into $\mathcal{D}$ intervals of length $\frac{T_{\max}}{\mathcal{D}}$,
each with a random population size $\frac1{\eta_d} \sim \text{logUniform}(0.1,10)$,
with 10 repetitions per setting of $T_{\max}$ and $\mathcal{D}$.
Empirically, LDpop scales sublinearly with $\mathcal{D}$,
and LDpop has no problem handling $\mathcal{D}=64$ epochs.
We also note that $T_{\max}$ has a greater impact on runtime than $\mathcal{D}$;
this is because the matrix exponentials
are essentially computed by solving an ODE from $-T_{\max}$ to $0$,
as noted in Appendix~\ref{appendix:action}.

\subsection{Accuracy of the approximate likelihood \label{sec:approx:accuracy}}

\begin{figure}[t]
    \centering
    \begin{subfigure}[t]{0.43\textwidth}
\includegraphics[width=\textwidth]{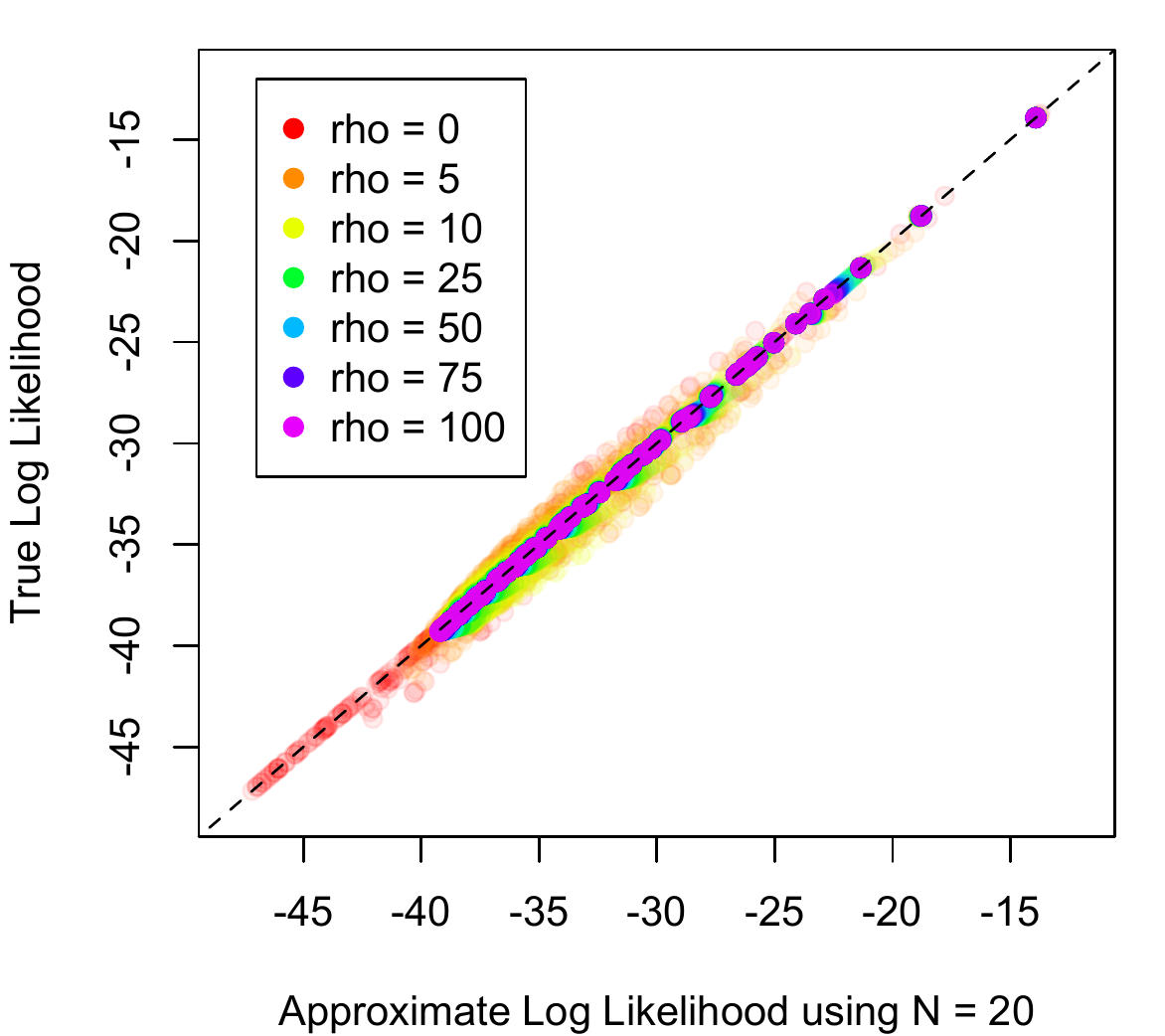}
        \caption{}
    \end{subfigure}
    \qquad \qquad 
    \begin{subfigure}[t]{0.43\textwidth}
\includegraphics[width=\textwidth]{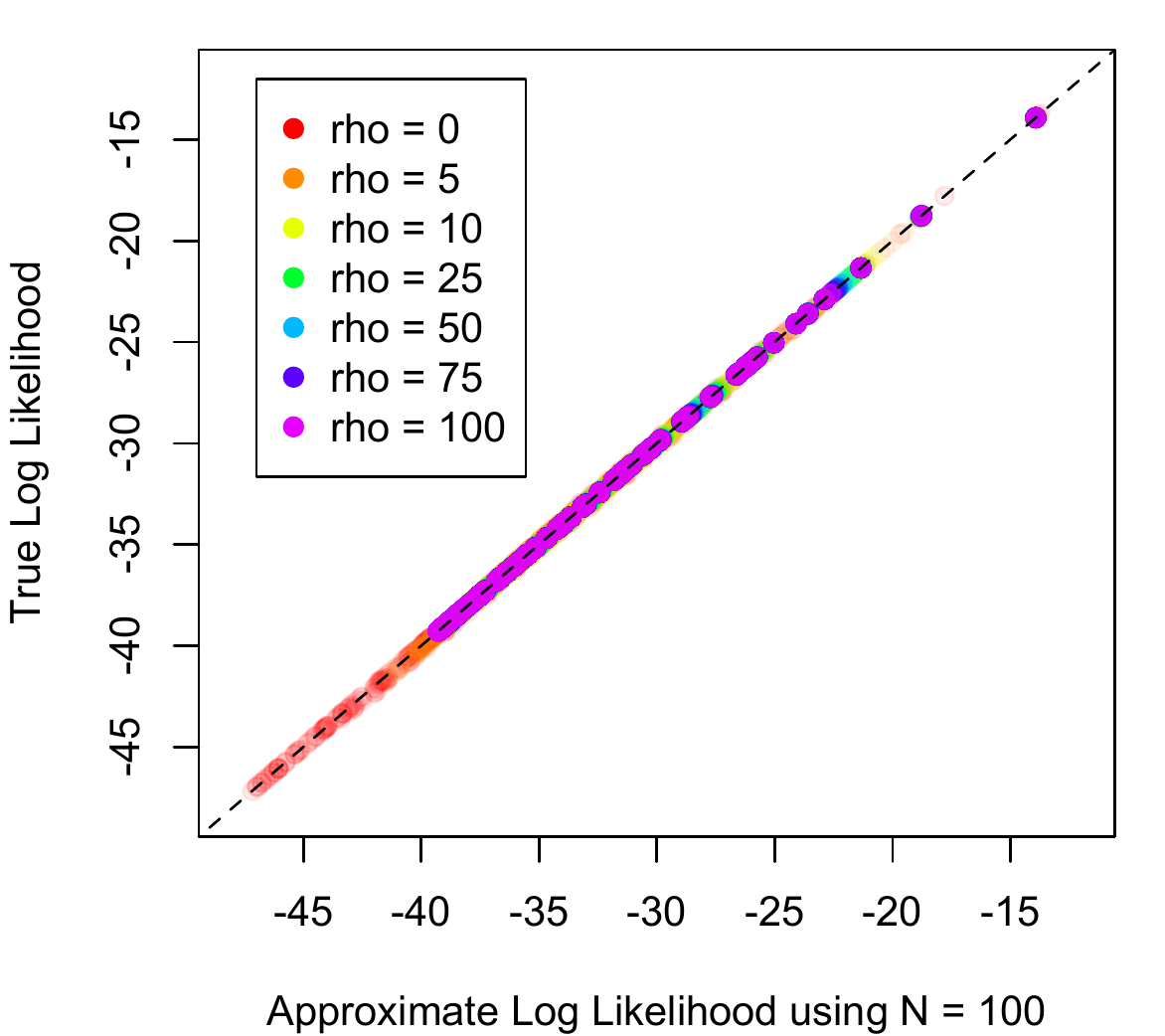}
        \caption{}
    \end{subfigure}
\caption{The approximate table $\{\log \hat{\mathbb{P}}(\mathbf{n})\}$ plotted against the exact table $\{\log \mathbb{P}(\mathbf{n})\}$, for a lookup table with $n=20$ and $\rho\in\{0,1,\ldots,100\}$, under the 3-epoch model in \eqref{eq:exampleDemo}. (a) $N=20$ Moran particles. (b) $N=100$ Moran particles. 
Note that the approximate table with $N=100$ is extremely accurate, and visually
indistinguishable from the true values.
}
\label{fig:ek:error}
\end{figure}

In Section~\ref{sec:ldhat} we found that using the approximate likelihood
\eqref{eq:moran} has little impact on recombination rate estimation,
suggesting that it is an accurate approximation to the exact formula
in Theorem~\ref{thm:augmented:moran}.

We examine this in greater detail in Figure~\ref{fig:ek:error},
for $n=20$, the 3-epoch demography \eqref{eq:exampleDemo}, and a lookup table
with $\rho \in \{0,1,\ldots,100\}$.
We compare the approximate against the exact values for $N=20$ and
$N=100$ Moran particles in the approximate model.
The approximate table with $N=20$ is reasonably accurate, with some mild
deviations from the truth.
The approximate table with $N=100$ is extremely accurate, and visually
indistinguishable from the true values.

\subsection{Runtime and accuracy of the importance sampler\label{sec:is:accuracy:runtime}}

\begin{figure}
  \centering
\begin{subfigure}[t]{.47\textwidth}
\begin{center}
\includegraphics[width=\textwidth]{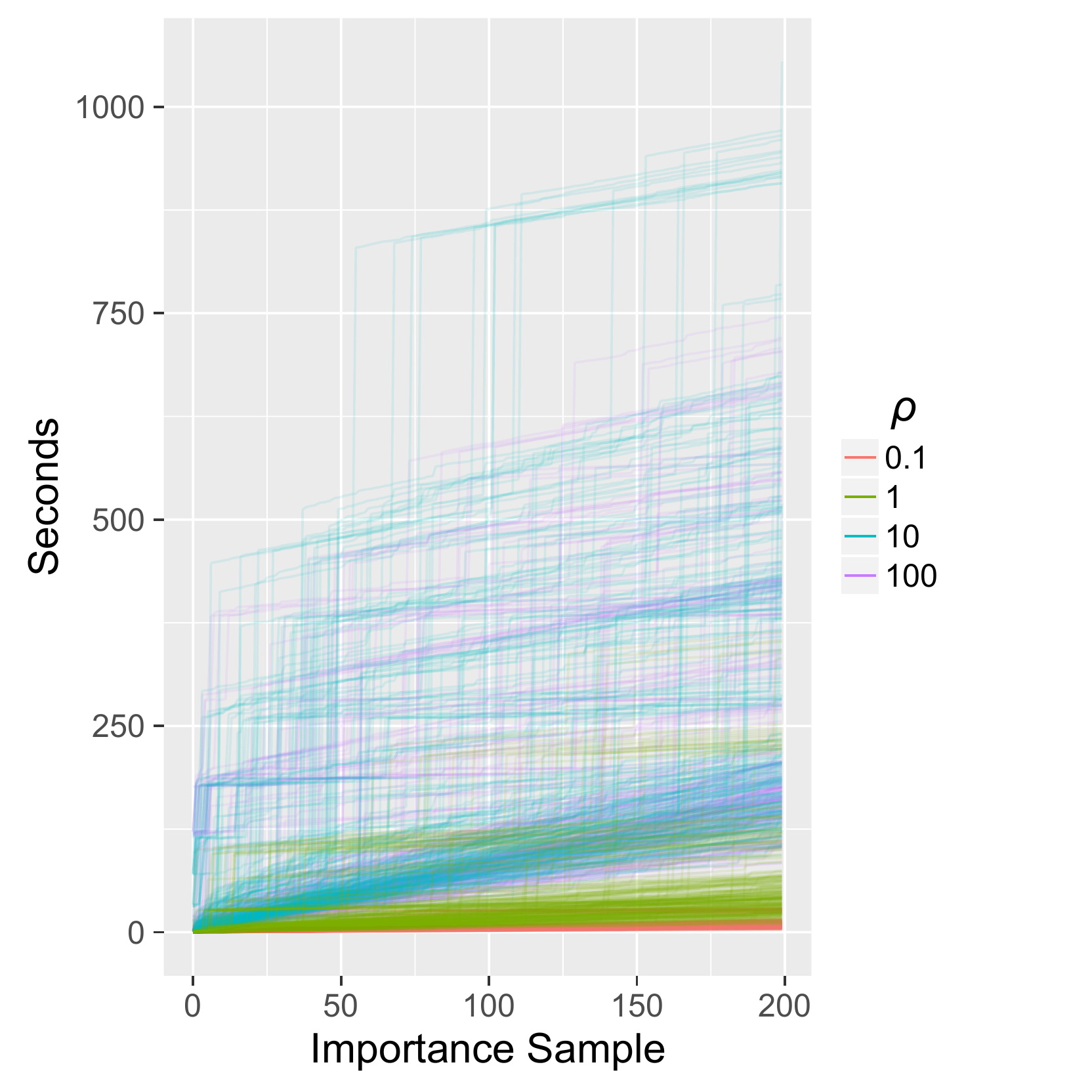}
\caption{}
\label{fig:is:time}
\end{center}
\end{subfigure}
\qquad
\begin{subfigure}[t]{.47\textwidth}
\begin{center}
\includegraphics[width=\textwidth]{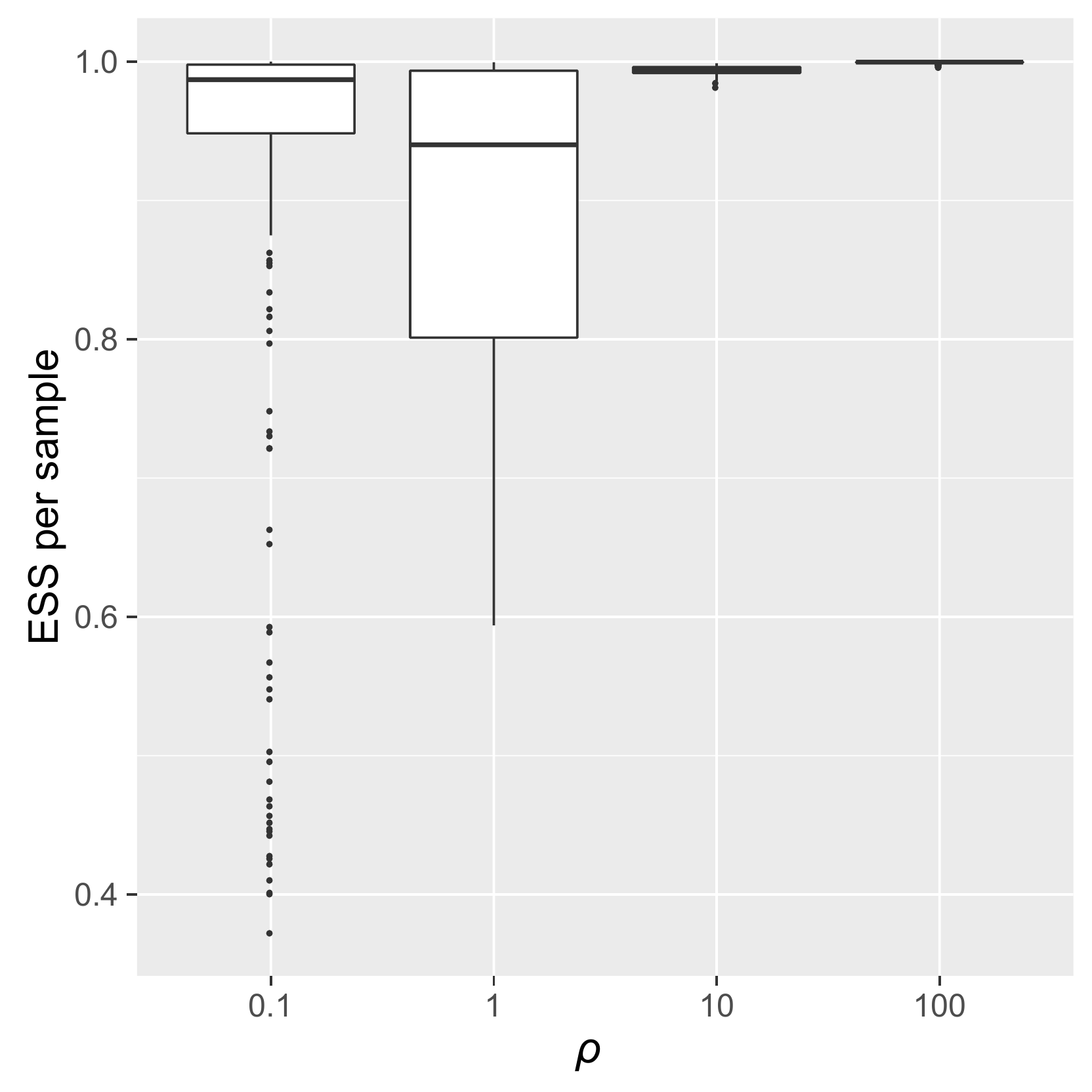}
\caption{}
\label{fig:ess}
\end{center}
\end{subfigure}
\begin{subfigure}[t]{.47\textwidth}
\begin{center}
\includegraphics[width=\textwidth]{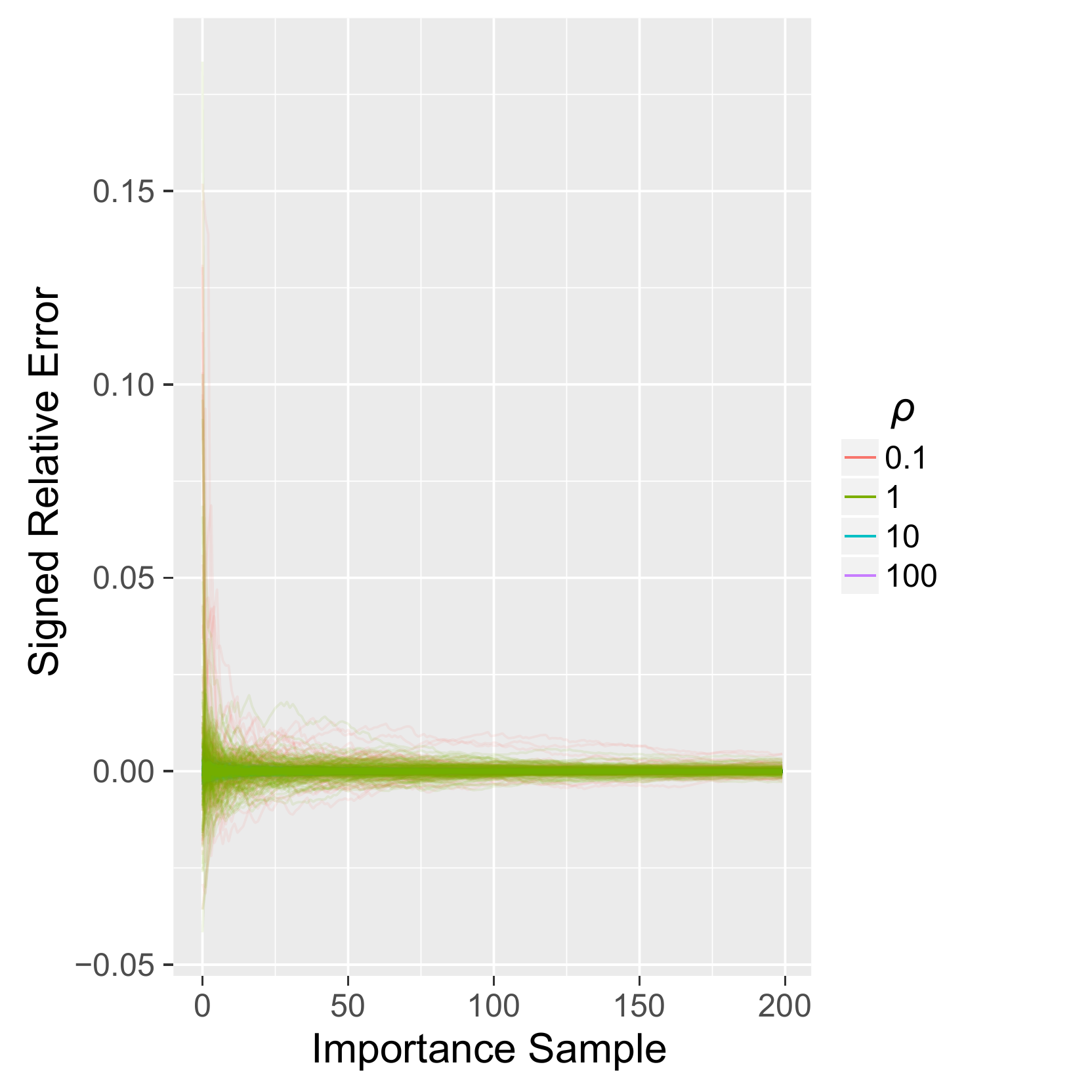}
\caption{}
\label{fig:comparison}
\end{center}
\end{subfigure}
  \caption{Accuracy and runtime of importance sampling, on the 3-epoch demography \eqref{eq:exampleDemo} with $\theta=0.008$ and $\rho=1.0$, drawing 200 genealogies for each of the 275 fully-specified configurations $\mathbf{n}$ with $n=20$. 
  (a) Runtime for each $\mathbf{n}$, as a function of the number of importance samples. Higher $\rho$ generally took a longer time. Using 20 cores, the time to sample all 275 configurations took about 4 minutes when $\rho=0.1$, but 1 hour for $\rho=100$.
  (b) The ESS per importance sample, for each configuration $\mathbf{n}$.
  (c) The signed relative error $\frac{\text{Est}-\text{Truth}}{\text{Truth}}$ of $\log \hat{\mathbb{P}}(\mathbf{n})$, as a function of the number of importance samples. The true values were computed via Theorem~\ref{thm:augmented:moran}.
  }
  \label{fig:is:results}
\end{figure}

We plot the runtime of the importance sampler (Section~\ref{sec:importance:sampling})
in Figure~\ref{fig:is:time}.
For the previous 3-epoch demography \eqref{eq:exampleDemo}, we drew
$K=200$ importance samples for each of the 275 configurations $\mathbf{n}$
with $n=n^{(c)}=20$, with $\rho \in \{0.1, 1, 10, 100\}$.
The runtime of the importance sampler generally increased with $\rho$; 
using 20 cores, sampling all 275 configurations took about 4 minutes with
$\rho=0.1$, but about 1 hour with $\rho=100$.
We further analyze the computational complexity of the importance sampler in
Appendix~\ref{appendix:is:complexity}.

The number  $K$  of importance samples required to reach
a desired level of accuracy is typically measured with
the effective sample size (ESS):
\begin{align*}
  \text{ESS} &= \frac{\left(\sum_{k=1}^K w_k\right)^2}{\sum_{k=1}^K w_k^2},
\end{align*}
where $w_k = \frac{d\mathbb{P}(\mathbf{n}^{(k)}_{\leq 0})}{d \hat{\proposal}(\mathbf{n}^{(k)}_{\leq 0})}$ denotes the importance weight of the $k$th sample (see \eqref{eq:importance:sampling}).
Note that $\text{ESS} \leq K$ always, with equality only achieved
if the $w_k$ have $0$ variance.

Compared to previous coalescent importance samplers, our
proposal distribution is highly efficient.
We plot the ESS per importance sample (i.e. $\frac{\text{ESS}}K$)
in Figure~\ref{fig:ess}.
Typically $\frac{\text{ESS}}K > 0.8$; for $\rho\in\{10,100\}$,
the ESS is close to its optimal value, with $\frac{\text{ESS}}K \approx 1$.
In Figure~\ref{fig:comparison}, we compare the log-likelihood estimated
from importance sampling to the true value computed with Theorem~\ref{thm:augmented:moran};
after $K=200$ importance samples, the signed relative error is well under $1\%$ for all $\mathbf{n}$.

By contrast, the previous two-locus importance sampler of \citet{fearnhead2001estimating}, which assumes a constant population size,
achieves ESS anywhere between $0.05K$ and $0.5K$,
depending on $\mathbf{n}, \theta, \rho$  (result not shown).
This importance sampler is based on
a similar result as Theorem~\ref{thm:posterior:chain},
with optimal rates $\isQ_{\mathbf{n},\mathbf{m}} \frac{\mathbb{P}(\mathbf{m})}{\mathbb{P}(\mathbf{n})}$.
However, to approximate $\frac{\mathbb{P}(\mathbf{m})}{\mathbb{P}(\mathbf{n})}$,
previous approaches did not use a Moran model,
but followed the approach of \citet{stephens2000inference},
using an approximate ``conditional sampling distribution'' (CSD).
We initially tried using the CSD of \citet{fearnhead2001estimating}
and later generalizations to variable demography \citep{sheehan2013estimating, steinruecken2015inference},
but found that importance sampling failed under population bottleneck scenarios,
with the ESS repeatedly crashing to lower and lower values.
Previous attempts to perform importance sampling under variable demography \citep{ye2013importance}
have also encountered low ESS,
though in the context of an infinite sites model (as opposed to a 2-locus model).
However, \citet{dialdestoro2016coalescent} recently devised an efficient 2-locus
importance sampler using the CSD-approach, in conjunction with advanced importance sampling
techniques. Their importance sampler allows archaic samples and is thus time-inhomogeneous,
but it only models constant population size histories.

\section{Discussion}

In this paper, we have developed a novel algorithm for computing the exact two-locus sampling probability under arbitrary piecewise-constant demographic histories.  These two-locus likelihoods can be used to study the impact of demography on LD and also to improve fine-scale recombination rate estimation.  Indeed, using two-locus sampling probabilities computed under the true or an inferred demography, we were able to obtain recombination rate estimates with substantially less noise and fewer spurious peaks that could potentially be mistaken for hotspots.  

We have implemented our method in a freely available software package, LDpop.  This program also includes an efficient approximation to the true sampling probability that easily scales to hundreds in sample size.   In practice, highly accurate approximations to the true sampling probability for sample size $n$ can be obtained quickly by first applying the approximate algorithm with $N$ Moran particles larger than $n$ and then down-sampling to the desired sample size $n$. 

In principle, one could also obtain an accurate approximation to the sampling probability using our importance sampler, which is also implemented in LDpop.  We have not optimized this code, however, and we believe that its main utility will be in sampling two-locus ARGs from the posterior distribution.  Lastly, we note that, in addition to improving the inference of fine-scale recombination rate variation, our two-locus likelihoods can be utilized in other applications such as hotspot hypothesis testing and demographic inference.

\section*{Acknowledgments}
This research is supported in part by an NIH grant R01-GM108805, an NIH training grant T32-HG000047, and a Packard Fellowship for Science and Engineering.

\clearpage
\appendix
\section*{Appendix}

\section{Proposal distribution of importance sampler \label{appendix:proposal}}

For the importance sampler of Section~\ref{sec:importance:sampling},
we construct the proposal distribution
$\hat{\proposal}(\mathbf{n}_{\leq 0})$
by approximating the optimal proposal distribution
$\proposal_{\text{opt}}(\mathbf{n}_{\leq 0}) =
\mathbb{P}(\mathbf{n}_{\leq 0} \mid \mathbf{n}_0)$ given in Theorem~\ref{thm:posterior:chain}.
We start by choosing a grid of points $-\infty < \tau_1 < \tau_2 < \cdots < \tau_J = 0$,
then set $\hat{\proposal}$ to be a backwards in time Markov chain,
whose rates at $t \in (\tau_j, \tau_{j+1})$ are the linear interpolation
\begin{align}
  \hat{\isRate}^{(t)}_{\mathbf{n},\mathbf{m}} &= 
\frac{\tau_{j+1} - t}{\tau_{j+1} - \tau_j}
\hat{\isRate}^{(\tau_j)}_{\mathbf{n},\mathbf{m}}
+
\frac{t - \tau_j}{\tau_{j+1} - \tau_j}
\hat{\isRate}^{(\tau_{j+1})}_{\mathbf{n},\mathbf{m}},
\label{eq:interpolation}
\end{align}
with the rates at the grid points given by
\begin{eqnarray*}
  \hat{\isRate}^{(\tau_j)}_{\mathbf{n}, \mathbf{m}}
&=&
\begin{cases}
\isQ^{(\tau_j)}_{\mathbf{n},\mathbf{m}}
\frac{\mathbb{\hat{P}}_{\tau_j}(\mathbf{m})}{\mathbb{\hat{P}}_{\tau_j}(\mathbf{n})},
& \text{if } \mathbf{m} \neq \mathbf{n},
\\
-\sum_{\bfmath{\nu} \neq \mathbf{n}} \hat{\isRate}^{(\tau_j)}_{\mathbf{n}, \bfmath{\nu}},
& \text{if } \mathbf{m} = \mathbf{n},
\end{cases}
\end{eqnarray*}
with $\mathbb{\hat{P}}_{\tau_j}(\mathbf{n})$ an approximation 
to the likelihood $\mathbb{P}_{\tau_j}(\mathbf{n})$. 
In particular, we
set $\mathbb{\hat{P}}_{\tau_j}(\mathbf{n}) = \mathbb{P}^{(N)}_{\tau_j}(\mathbf{n})$, 
using the approximate likelihood formula \eqref{eq:moran}
in Section~\ref{sec:moran:dk}.
The approximate likelihoods $\{\mathbb{P}^{(N)}_{\tau_j}(\mathbf{n})\}$ can be efficiently
computed along a grid of points using the method of Appendix~\ref{appendix:action} (see also
Appendix~\ref{appendix:is:complexity}).

To sample from $\hat{\proposal}$, 
we note that for configuration $\mathbf{n}$ at time $t$, the time $S < t$ of the next event has CDF
$\mathbb{P}(S < s) = \exp(\int_S^t \hat{\isRate}_{\mathbf{n},\mathbf{n}}^{(u)} du)$ for $s < t$.  Thus,
$S$ can be sampled by first sampling $X \sim \text{Uniform}(0,1)$, and then solving for
$\log(X) = \int_S^t \hat{\isRate}_{\mathbf{n},\mathbf{n}}^{(u)} du$
via the quadratic formula (since $\hat{\isRate}_{\mathbf{n},\mathbf{n}}^{(u)}$
is piecewise linear; see \eqref{eq:interpolation}).
Having sampled $S$, we can then sample the next configuration $\mathbf{m}$
with probability $-{\hat{\isRate}^{(S)}_{\mathbf{n},\mathbf{m}}}/{\hat{\isRate}^{(S)}_{\mathbf{n},\mathbf{n}}}$.

\section{Details of simulation study in Section~\ref{sec:ldhat} \label{appendix:details}}

\subsection{Simulated data \label{appendix:simulated:details}}

We simulated independent 1 Mb segments with $n = 20$ haplotypes under the 3-epoch demography $\eta(t)$ in \eqref{eq:exampleDemo}.
To do so, we generated trees using the program MaCS \citep{chen2009fast}, and then generated mutations according to a quadra-allelic mutational model.
For the variable recombination maps used in Figures~\ref{fig:representative:rho} and \ref{fig:r2:all},
we divided the recombination map of the X chromosome of \emph{D. melanogaster} from Raleigh, NC inferred by \citet{chan2012genome} into 22 non-overlapping 1~Mb windows and simulated 5 replicates, for a total of 110~Mb of simulated data.
For the constant map used in Figure~\ref{fig:constant:rho}, we generated 20 datasets with $\rho=0.01$ per base.

\subsection{Estimation of misspecified demography $\hat{\eta}(t)$ \label{appendix:misspecified}}

To estimate the misspecified demography $\hat{\eta}(t)$ of \eqref{eq:missDemo},
we pooled all bi-allelic SNPs from the 110 simulated segments of the variable recombination map,
and then used the folded site frequency spectrum (SFS) of the simulated SNPs to estimate
$\hat{\eta}(t)$.
Specifically, we fit $\hat{\eta}(t)$ by maximizing a composite likelihood,
viewing each SNP as an independent draw from a multinomial distribution
proportional to the expected SFS.
We computed the expected SFS with the software package \texttt{momi} \citep{kamm2016efficient}, and fixed the most ancient population size to $1.0$ due to scaling and identifiability issues.

\subsection{Recombination map estimation}

After removing all non-biallelic SNPs, we ran both LDhat and LDhelmet on the resulting data, 
using a block penalty of 50 as recommended by \citet{chan2012genome} for \emph{Drosophila}-like data
(the block penalty is a tuning parameter that is multiplied by the number of change-points in the estimated map $\bfmath{\hat{\rho}}$, and added to the log composite likelihood; thus a high block penalty discourages over-fitting).
We took the posterior median inferred at each position to be the estimated map $\bfmath{\hat{\rho}}$.  We only used the centermost 500~kb of each estimate $\bfmath{\hat{\rho}}$ to avoid the issue of edge effects.

\section{Computational complexity\label{appendix:complexity}}

\subsection{Computing the action of a sparse matrix exponential \label{appendix:action}}

Both Theorem~\ref{thm:augmented:moran} and the approximate formula \eqref{eq:moran}
rely on ``the action of the matrix exponential'' \citep{al-mohy2011computing}.
Let $\bfA$ be a $k \times k$ matrix and $\bfv$ a $1 \times k$ row vector.
We need to compute expressions of the form $\bfv e^\bfA$.
Naively, this kind of vector-matrix multiplication costs $O(k^2)$.
However, in our case $\bfA$ will be sparse, with $k$ nonzero entries,
allowing us to more efficiently compute $\bfv e^\bfA$.

In particular, we use the algorithm of \citet{al-mohy2011computing},
as implemented in the Python package \texttt{scipy}.
For $s \in \mathbb{Z}_+$, define $T_m(s^{-1} \bfA) = \sum_{i=0}^m {(s^{-1} \bfA)^i}/{i!}$,
the truncated Taylor series approximation of $e^{s^{-1} \bfA}$. Then, we have
\begin{align*}
  \bfv e^\bfA &= \bfv \left(e^{s^{-1} \bfA} \right)^s \approx \bfv [T_m(s^{-1} \bfA)]^s.
\end{align*}
Now let $\bfb_j = \bfv [T_m(s^{-1} \bfA)]^j$, so $B_j$ is a $1 \times k$ row vector. Then
\begin{align*}
  \bfb_j &= \bfb_{j-1} T_m(s^{-1} \bfA) = \sum_{i=0}^m \bfb_{j-1} \frac{(s^{-1}\bfA)^i}{i!},
\end{align*}
with $\bfv e^\bfA \approx \bfb_s$, and $\bfb_s$ evaluated in $\mathcal{T} = ms$ vector-matrix
multiplications, each costing $O(k)$ by the sparsity of $\bfA$.
Approximating $\bfv e^\bfA$ thus costs $O( \mathcal{T} k)$ time.
Both $m,s$ are chosen automatically  to bound
\begin{align*}
  \frac{\| \Delta \bfA\|_1}{\|\bfA\|_1} &\leq \text{tolerance} \approx 1.1 \times 10^{-16},
\end{align*}
with $\Delta \bfA$ defined by $[T_m(s^{-1} \bfA)]^s = e^{\bfA + \Delta \bfA}$,
and the matrix norm given by $\|\bfA\|_1 = \sup_{\mathbf{w} \ne \mathbf{0}} \frac{\|\mathbf{w} \bfA\|_1}{\|\mathbf{w}\|_1}$.
To avoid numerical instability, $m$ is also bounded by
$m \leq m_{\max} = 55$.
\citet{al-mohy2011computing} provide some analysis for the size of $\mathcal{T} = m s$,
but this analysis is rather involved. Very roughly, $\mathcal{T}$ is proportional to
$\|\bfA\|$ (for arbitrary matrix norm $\| \cdot \|$), so computing $\bfv e^{2\bfA}$ takes twice as long as $\bfv e^{\bfA}$, and computing $\bfv e^{t \bfA}$ is roughly proportional to $t$.
This is because $\bfv e^{t \bfA}$ is essentially computed by numerically integrating the ODE $\nabla \bfmath{f}(s) = \bfmath{f}(s) \bfA$ for $s \in [0,t]$.

We note that $\bfb_j \approx \bfv e^{s^{-1} j \bfA}$, 
and thus this algorithm approximates $\bfv e^{t \bfA}$
along a grid of points $t \in \{s^{-1}, 2s^{-1}, \ldots, 1\}$.
If $\bfv e^{t \bfA}$ is needed at additional points,
then extra grid points can be added at those times.

\subsection{Complexity of the exact likelihood formula (Theorem~\ref{thm:augmented:moran})\label{appendix:exact:complexity}}

We consider the computational complexity of computing $\mathbb{P}_0(\mathbf{n})$ via Theorem~\ref{thm:augmented:moran}.
Note that the formula \eqref{eq:thm1} simultaneously computes $\mathbb{P}_0(\mathbf{n})$ for all configurations $\mathbf{n} \in \mathcal{N}$.

As usual, we assume 2 alleles $\mathcal{A}=\{0,1\}$, as is assumed by LDhat and the applications in Section~\ref{sec:empirical}.
We start by considering the dimensions of the sampling probability vectors $\bfp^\timeInd$ and rate matrices $\tilde{\Rate}^\timeInd$ for intervals $(t_\timeInd,t_{\timeInd+1}]$. The set of $a,b,c$ haplotypes is $H = \{00,01,10,11,0*,1*,*0,*1\}$, so $|H| = 8$.  Thus, there are $O(n^6)$ possible configurations $\mathbf{n}$ with $n^{(a)} = n^{(b)} = n - n^{(c)}$. In particular, there are $O(n^6)$ ways to specify $n_{00}, n_{01}, n_{10}, n_{11}, n_{0*}, n_{*0}$, and then $n_{1*} = n-\sum_{i,j \in \{0,1\}} n_{ij} - n_{0*}$ and $n_{*1} = n-\sum_{i,j \in \{0,1\}} n_{ij} - n_{*0}$ are determined.
Thus, $\bfp^\timeInd$ is a row vector of dimension $1 \times O(n^6)$, and $\tilde{\Rate}^\timeInd$ is a square matrix of dimension $O(n^6) \times O(n^6)$, but $\tilde{\Rate}^\timeInd$ is sparse, with only $O(n^6)$ nonzero entries.

By using the algorithm of Appendix~\ref{appendix:action},
we can compute $\bfp^{\timeInd+1} = \big[(\bfp^{\timeInd} \odot \tilde{\stationaryGamma}^{\timeInd})
e^{\tilde{\Rate}^\timeInd (t_{\timeInd+1} - t_\timeInd)}\big] \div \tilde{\stationaryGamma}^{\timeInd}$
from $\bfp^\timeInd$
in $O(\mathcal{T}_{\timeInd} n^6)$ time,
where $\mathcal{T}_{\timeInd}$ is the number of vector-matrix multiplications
to compute the action of $e^{\tilde{\Rate}^\timeInd (t_{\timeInd+1} - t_\timeInd)}$.
We note that the stationary distribution $(\gamma^\timeInd_0,\ldots,\gamma^\timeInd_n)$
can be computed in $n+1$ steps:
${\bfGamma}^\timeInd$ is the rate matrix of a simple random walk with $n+1$ states,
so $\gamma^{\timeInd}_{i+1} = \gamma^\timeInd_i \frac{[{\Gamma}^d]_{i,i+1}}{[{\Gamma}^d]_{i+1,i}}$
and $\sum_{i} \gamma^{\timeInd}_{i} = 1$.

Similarly, the initial value
$\bfp^{-\pieces+1} = \stationaryLambdaTilde{-\pieces} \div \tilde{\stationaryGamma}^{-\pieces}$ can be computed
via sparse
vector-matrix multiplications, using the technique of power iteration.
For $\mu = \frac1{\max_{ij} [\tilde{\Rate}^{-\pieces}]_{ij}}$ and
arbitrary positive vector $\bfv^{(0)}$ with $\|\bfv^{(0)}\|_1 = 1$,
we have $\bfv^{(i)} := \bfv^{(0)} (\mu \tilde{\Rate}^{-\pieces} + I)^{i} \to \stationaryLambdaTilde{-\pieces}$ as $i \to \infty$.
In particular, we set the number of iterations, $\mathcal{T}_{-\pieces}$, so that
$\left\| \log (\bfv^{(\mathcal{T}_{-\pieces})} \div \bfv^{(\mathcal{T}_{-\pieces}-1)}) \right\|_1 < 1 \times 10^{-8}$,
where $\log \bfv$ is the element-wise $\log$ of $\bfv$.
As noted in Section~\ref{sec:runtime:exact:approx},
in practice we found $\mathcal{T}_{-\pieces} \gg \mathcal{T}_d$ for $d > -\pieces$,
i.e. computing the initial stationary distribution $\stationaryLambdaTilde{-\pieces}$ was more expensive
than multiplying the matrix exponentials $e^{\tilde{\Rate}^\timeInd t}$.

To summarize, computing $\mathbb{P}_0(\mathbf{n})$ for all $O(n^6)$ configurations $\mathbf{n}\in\mathcal{N}$ of size $n$ costs $O(n^6 \mathcal{T})$,
with $\mathcal{T} = \sum_{d=-\pieces}^{-1} \mathcal{T}_{d}$.
We caution that $\mathcal{T}$ depends on
$n, \{t_\timeInd\}, \{\tilde{\Rate}^\timeInd\}$.

The memory cost of Theorem~\ref{thm:augmented:moran} is $O(n^6)$, since
$\tilde{\Rate}^\timeInd$ has $O(n^6)$ nonzero entries.

\subsubsection{Comparison with Golding's equations}

Under constant population size,
\citet{golding1984sampling} proposed a method to compute $\mathbb{P}_0(\mathbf{n})$
by solving a linear system of equations $\mathbf{g} \mathbf{G} = \mathbf{g}$,
where $\mathbf{g} = [\mathbb{P}_0(\mathbf{n})]_{\mathbf{n} \in \mathcal{N}'}$ is the vector
of sampling probabilities indexed by the $O(n^8)$ configurations $\mathcal{N}' = \{\mathbf{n}: \max(n^{(a)}+n^{(c)},n^{(b)}+n^{(c)}) \leq n\}$
with at most $n$ alleles at each locus.
\citet{hudson2001two} solves this linear system, costing $O(n^8 \mathcal{T})$ where (as above) $\mathcal{T}$ is some finite number of
sparse matrix-vector multiplications.

For the case of constant population size, Theorem~\ref{thm:augmented:moran}
reduces to solving a sparse system $\stationaryLambdaTilde{-D} \tilde{\Rate}^{-D} = \stationaryLambdaTilde{-D}$,
which is similar in spirit to solving Golding's equations $\mathbf{g} \mathbf{G} = \mathbf{g}$.
The $O(n^6 \mathcal{T})$ runtime of Theorem~\ref{thm:augmented:moran} at first seems superior
to the $O(n^8 \mathcal{T})$ runtime of Golding's equations,
but in fact the number of matrix multiplications $\mathcal{T}$ is not comparable between the two methods.
Most importantly, \citet{hudson2001two} exploits the structure of the $O(n^8)$ equations to decompose them into smaller sub-systems of $O(n^4)$
equations, which may lead to smaller $\mathcal{T}$.
Algorithmic details also lead to important differences: we use power iteration, whereas \citet{hudson2001two}
use a conjugate gradient method, with less stringent convergence criteria (stopping when the relative $l_2$ error
for each sub-system of equations is $< 10^{-4}$).

Preliminary tests suggest that the C code of \citet{hudson2001two}, and a similar implementation by \citet{chan2012genome},
are faster than our current method for solving $\stationaryLambdaTilde{-D} \tilde{\Rate}^{-D} = \stationaryLambdaTilde{-D}$.
We are planning future updates to LDpop that will speed up the initial stationary distribution $\stationaryLambdaTilde{-D}$,
either by changing the algorithmic details of our solver, or by using Golding's equations to compute $\stationaryLambdaTilde{-D}$ instead.

\subsection{Complexity of approximate likelihood formula}

The method of computing the approximate likelihood formula \eqref{eq:moran}
is similar to computing Theorem~\ref{thm:augmented:moran},
in that we can compute an initial stationary distribution $\stationaryLambda^{-\pieces}_{(N)}$
by power iteration, and then propagate it forward in time by applying the
action of the sparse matrix exponential $e^{\Rate_{(N)}^\timeInd (t_{\timeInd+1} - t_\timeInd)}$.
However, instead of $O(n^6)$ states, there are $O(N^3)$ total states:
there are 4 possible fully-specified haplotypes $\{00, 01, 10, 11\}$,
and thus the requirement that the number of lineages sums to $N$ yields $O(N^3)$ possible states
for the Moran model $\mathbf{M}_t$.
Thus, computing the approximate likelihood formula \eqref{eq:moran}
costs $O(N^3 \mathcal{T})$ time and $O(N^3)$ memory space.

\subsection{Complexity of importance sampler\label{appendix:is:complexity}}

Here we examine the computational complexity of the importance sampler of Section~\ref{sec:importance:sampling}.

To construct the proposal distribution $\hat{\proposal}$ (Appendix~\ref{appendix:proposal}),
we must compute approximate likelihoods $\mathbb{P}^{(N)}_t(\mathbf{n})$
defined by \eqref{eq:moran}
along a grid of points $t \in \{\tau_1, \tau_2, \ldots, \tau_J\}$.
We start by computing the Moran likelihoods $\{\mathbb{P}_{(N)}(\moran_{\tau_j})\}_j$
using the action of the sparse matrix exponential.
As discussed in Appendix~\ref{appendix:action},
the method of \citet{al-mohy2011computing}
yields $\{\mathbb{P}_{(N)}(\moran_{\tau_j})\}_j$ as a byproduct of
computing $\mathbb{P}_{(N)}(\moran_0)$ at essentially no extra cost.
Thus, computing the terms $\{\mathbb{P}_{(N)}(\moran_t)\}$ costs
$O(N^3 \mathcal{T})$ (absorbing the minor cost of an additional $J$ extra grid points
into $\mathcal{T}$).

We then compute $\mathbb{P}^{(N)}_t(\mathbf{n})$ by
subsampling from $\mathbb{P}_{(N)}(\moran_t)$ as in
\eqref{eq:moran}, and thus set $N = 2n$, since $2n$ is the maximum
number of individuals in $\mathbf{n}$ (because each of the original
$n$ lineages can recombine into two lineages).
However, it is inefficient to compute $\mathbb{\hat{P}}_t(\mathbf{n})$ by subsampling for every value
of $\mathbf{n}$ separately.  Instead, it is  better to use a dynamic program
$\mathbb{\hat{P}}_t(\mathbf{n}) = \sum_{\mathbf{m}} \mathbb{\hat{P}}_t(\mathbf{m}) \mathbb{P}(\mathbf{n} \mid \mathbf{m})$,
where the sum is over all configurations $\mathbf{m}$ obtained by adding an additional sample to $\mathbf{n}$.

This costs $O(n^8 J)$ time and space, since there are
$J$ grid points and $O(n^8)$ possible configurations of $\mathbf{n}$.
Then, assuming a reasonably efficient proposal, the expected cost to draw $K$ importance samples is $O(nJK)$,
since the expected number of coalescence, mutation, and
recombination events before reaching the marginal common ancestor at each locus is $O(n)$ \citep{griffiths1991two}.
This approach thus takes $O(n^3\mathcal{T} + n^8J + n^4JK)$ expected time
to compute $\mathbb{P}_0(\mathbf{n})$ for all $O(n^3)$ possible $\mathbf{n}$.
In practice,
we only precomputed $\mathbb{\hat{P}}_t(\mathbf{n})$ for the
$O(n^4)$ fully specified $\mathbf{n}$ (without missing alleles),
but computed and cached
$\mathbb{\hat{P}}_t(\mathbf{n})$
as needed
for partially specified $\mathbf{n}$ (with missing alleles).
The theoretical running time to compute the full lookup table is still
$O(n^3\mathcal{T}D + n^8J + n^4JK)$, but in practice,
many values of $\mathbf{n}$ are highly unlikely
and never encountered at each $\tau_j$.

\section{Proofs \label{sec:proofs}}

For a stochastic process $\{X_t\}_{t \leq 0}$, we denote its partial sample paths with the following notation:
$X_{s:t} = \{X_u : u \in (s,t]\}$ and $X_{\leq s} = X_{-\infty:s}$.

\subsection{Proof of Theorem~\ref{thm:augmented:moran} \label{sec:proof:augmented:moran}}

We start by constructing a \emph{forward-in-time} Markov jump process
$\moranModified_{\leq 0}$ with state space
$\mathcal{N} = \{\mathbf{n} : n^{(abc)} = (k,k,n-k), 0 \leq k \leq n\}$.
$\moranModified_t$ changes due to four types of events:
mutation, copying, ``recoalescence'', and ``recombination''.
\begin{enumerate}
\item Individual alleles mutate at rate
  $\frac{\theta}2$ according to transition matrix $\mathbf{P}$.
\item Lineages copy their alleles onto each other, with the rate
depending on the lineage type.
Each pair of $a$ types experiences a copying event at rate
$\frac1{\eta_\timeInd}$, with the direction of copying chosen
with probability $\frac12$.
The rates are the same for every pair of $b$ and
every pair of $c$ types.
Pairs of $(a,c)$ and $(b,c)$ types also experience
copying at rate $\frac1{\eta_\timeInd}$, however
the direction of copying is always from the $c$ type
to the $a$ or $b$ type, and only happens at one allele
(left for $a$, right for $b$).
\item $a$ types do not copy onto $b$ types, and vice versa.
Instead, they merge (``recoalesce'') into a single $c$ type at rate
$\frac1{\eta_\timeInd}$ per pair. Note this is similar to
the coalescent, but here the recoalescence
happens \emph{forward in time} rather than \emph{backwards in time}.
\item Each $c$ type splits into $a$ and $b$ types at rate
  $\frac{\rho}2$.
  Again, this is similar to the coalescent,
however here the ``recombination'' happens
\emph{forward in time}, while in the coalescent with recombination
it happens at rate $\frac{\rho}2$ going \emph{backwards in time}.
\end{enumerate}
Then $\moranModified_t$ has forward-in-time rate matrix $\Rate^\timeInd$ in $(t_\timeInd, t_{\timeInd+1}]$,
with $\Rate^\timeInd$ given in Table~\ref{tab:augmented:rates}.

Now let $C_t$ denote the number of $c$ types
in $\moranModified_t$ (so the number of $a$ and $b$ types are each $n-C_t$).
Note that $C_t$ is unaffected by mutation and copying events,
and so $C_{t+h}$ is conditionally independent of $\moranModified_t$
given $C_t$, for $h \geq 0$.
Thus $C_t$ is a Markov jump process with rate matrix $\bfGamma^\timeInd$ in $(t_\timeInd, t_{\timeInd+1}]$, 
where $\bfGamma^\timeInd$ is a tridiagonal square matrix indexed by $\{0,1,\ldots,n\}$,
with $\Gamma^d_{m,m-1} = \frac{\rho}2 m$, $\Gamma^d_{m,m+1} = (n-m)^2 \frac1{\eta_\timeInd}$, and $\Gamma^d_{m,m} = - \Gamma^d_{m,m-1} - \Gamma^d_{m,m+1}$.

We can therefore sample $\moranModified_{\leq 0}$ in two steps, as illustrated in Figure~\ref{fig:augmented:moran}:
\begin{enumerate}
\item First, sample the recoalescence and recombination events. In other words, sample $C_{\leq 0}$ using its rate matrices $\{\Gamma^d\}$.
\item Next, sample from $\mathbb{P}(\moranModified_{\leq 0} \mid C_{\leq 0})$.
For $h > 0$, $\moranModified_{t+h}$ can be obtained from $\moranModified_t$ and
$C_{\leq 0}$ by superimposing
two Poisson point processes, conditionally independent given $C_{\leq 0}$:
\begin{enumerate}
\item a point process of directed edges between lineages (the copying events),
  with rate $\frac1{2\eta_d}$ for $a \to a$, $b \to b$, $c \to c$ edges,
  and rate $\frac1{\eta_d}$ for $c \to a$, $c \to b$ edges.
\item a point process of mutations hitting the lineages,
  at rate $\frac{\theta}2$ per locus per lineage.
\end{enumerate}
To see that this superpositioning of point processes yields the correct distribution $\mathbb{P}(\moranModified_{\leq 0})$, note that mutation and copying events do not affect the rates of recombination and recoalescence events,
and that the 4 types of jump events that make up the Markov jump process $\moranModified_{\leq 0}$ occur
at the desired rates given by $\{\Rate^d\}$.
\end{enumerate}

\begin{figure}
  \centering
  \includegraphics[width=0.6\textwidth]{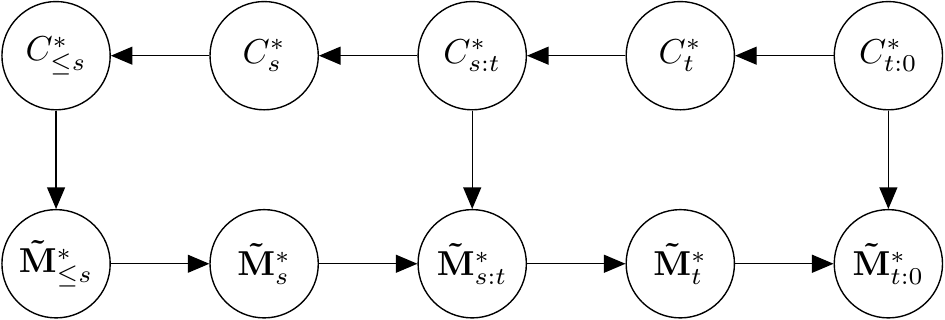}
  \caption{Probabilistic graphical model for the processes $C^*_t$ and $\moranModified^*_t$, with $-\infty < s < t \leq 0$.
Random variables are represented as vertices, and the edges encode conditional independence relationships. Specifically, if all paths between $V_1$ and $V_2$ pass through $W$, then there is conditional independence $\mathbb{P}(V_1, V_2 \mid W) = \mathbb{P}(V_1 \mid W) \mathbb{P}(V_2 \mid W)$.}
  \label{fig:augmented:graphical:model}
\end{figure}

Now define $C^*_t$ to be the \emph{backwards in time} Markov chain with
rates $\bfGamma^\timeInd$ in $(t_{\timeInd}, t_{\timeInd+1}]$ (whereas $C_t$ has the same rates but going \emph{forwards in time}).
Let $\moranModified^*_t$ be the stochastic process
with conditional law $\mathbb{P}(\moranModified^*_{\leq 0} \mid C^*_{\leq 0}=\mathcal{C}) = \mathbb{P}(\moranModified_{\leq 0} \mid C_{\leq 0}=\mathcal{C})$.
Thus $\moranModified^*_{\leq 0}$ can be sampled in the same two steps as
$\moranModified_{\leq 0}$, except the first step (coalescences and recombinations) is backwards in time.
We illustrate the conditional independence structure of $\moranModified^*_t$ and $C^*_t$ via a directed graphical model \citep{koller2009probabilistic} in Figure~\ref{fig:augmented:graphical:model}.
(A graphical model is a graph whose vertices represent random variables, with the property that if all paths between $V_1$ and $V_2$ pass through $W$, then there is conditional independence $\mathbb{P}(V_1, V_2 \mid W) = \mathbb{P}(V_1 \mid W) \mathbb{P}(V_2 \mid W)$).

We next show that for $\mathbf{n}$ with ${n}^{(a)} = {n}^{(b)} = n - {n}^{(c)}$,
\begin{eqnarray}
  \mathbb{P}_t(\mathbf{n})
&=&
\mathbb{P}(\moranModified^*_t = \mathbf{n} \mid C^*_t = n^{(c)})
.
\label{eq:forward:backward:moran}
\end{eqnarray}

We use a similar argument as in \citet[Theorem 1.30, p.47]{durrett2008probability},
tracing the genealogy of $\mathbf{n}$ backwards
in time (Figure~\ref{fig:augmented:moran:2}).
Under $\mathbb{P}(\moranModified^*)$, recombination events occur backwards in time
at rate $\frac{\rho}2$ per $c$ type lineage,
as in the coalescent. Likewise, coalescence between an $a$ and $b$
type occurs at the usual rate $\frac1{\eta_d}$.
Next, note that copying events between ancestral lineages induce coalescences within the ARG;
these are encountered as a Poisson point process at rate $\frac1{\eta_d}$ per ancestral
pair not of type $(a,b)$.
Thus, the embedded ARG is distributed as the coalescent with recombination.
Finally, conditioning on the full history of recombination, copying, and coalescence
events, we can drop down mutations as a Poisson point process with rate
$\frac{\theta}2$ per locus per lineage, and so the ARG with mutations follows
the coalescent with recombination and mutation.
Note that the alleles at the common ancestors of each locus
follow the stationary distribution:
the common ancestors are fixed under the conditioning (of recombination, copying, and coalescence events),
and if $\mathbf{v}_s$ is the conditional distribution of an ancestral allele at time
$s \leq T_{\text{MRCA}}$, then 
$\mathbf{v}_s = \mathbf{v}_{s'} e^{(\bfP - \mathbf{I}) \frac{\theta}2 (s - s')}$ for $s' < s$;
sending $s' \to -\infty$ yields the stationary distribution.

Having established \eqref{eq:forward:backward:moran},
we next observe
\begin{eqnarray}
  \mathbb{P}_{t_{\timeInd+1}}(\mathbf{n})
&=&
\mathbb{P}(\moranModified_{t_{\timeInd+1}}^* = \mathbf{n}
\mid C^*_{t_{\timeInd+1}}= n^{(c)})
\notag \\
&=&
\sum_{\mathbf{m}} \mathbb{P} (\moranModified_{t_{\timeInd}}^* = \mathbf{m} \mid C^*_{t_{\timeInd}} = m^{(c)})
\mathbb{P}(C^*_{t_{\timeInd}} = m^{(c)} \mid C^*_{t_{\timeInd+1}} = n^{(c)})
\notag \\
&& \qquad  \times \qquad
\mathbb{P}(\moranModified_{t_{\timeInd+1}}^* = \mathbf{n}
\mid C^*_{t_{\timeInd+1}}= n^{(c)}, C^*_{t_{\timeInd}} = m^{(c)}, \moranModified_{t_{\timeInd}}^* = \mathbf{m})
\notag \\
&=&
\sum_{\mathbf{m}} \mathbb{P}_{t_{\timeInd}} ( \mathbf{m})
\mathbb{P}(C^*_{t_{\timeInd}} = m^{(c)} \mid C^*_{t_{\timeInd+1}} = n^{(c)})
\notag \\
&& \qquad  \times \qquad
\mathbb{P}(\moranModified_{t_{\timeInd+1}} = \mathbf{n}
\mid C_{t_{\timeInd+1}}= n^{(c)}, C_{t_{\timeInd}} = m^{(c)}, \moranModified_{t_{\timeInd}} = \mathbf{m}).
\label{eq:forward:backward:moran:2}
\end{eqnarray}
Note that in the second equality, we use
the conditional independence of
$\moranModified_{t_{\timeInd}}^*$ and $C^*_{t_{\timeInd+1}}$ given
$C^*_{t_{\timeInd}}$, which follows from the graphical model of Figure~\ref{fig:augmented:graphical:model} by setting $s = t_\timeInd$ and $t = t_{\timeInd+1}$.

Next, note that $\bfGamma^\timeInd$ is the transition matrix of a
simple random walk with bounded state space and no absorbing states,
and thus is reversible.
Thus, with $\bfmath{\gamma}^\timeInd$ the stationary distribution of $\bfGamma^\timeInd$,
\begin{eqnarray}
\gamma^\timeInd_{n^{(c)}}
\mathbb{P}(C^*_{t_{\timeInd}} = m^{(c)} \mid C^*_{t_{\timeInd+1}} = n^{(c)})
&=&
\gamma^\timeInd_{n^{(c)}}
\Big[e^{\Gamma^{\timeInd} (t_{\timeInd+1} - t_{\timeInd})}\Big]_{n^{(c)},m^{(c)}}
\notag \\
&=&
\gamma^\timeInd_{m^{(c)}}
\Big[e^{\Gamma^{\timeInd} (t_{\timeInd+1} - t_{\timeInd})}\Big]_{m^{(c)},n^{(c)}}
\notag \\
&=&
\gamma^\timeInd_{m^{(c)}}
\mathbb{P}(C_{t_{\timeInd+1}} = n^{(c)}
\mid C_{t_{\timeInd}} = m^{(c)}).
\label{eq:moran:reversibility}
\end{eqnarray}
Recall that we defined $\tilde{\gamma}^\timeInd_{\bfn} = \gamma^\timeInd_{n^{(c)}}$. So plugging \eqref{eq:moran:reversibility} into \eqref{eq:forward:backward:moran:2} yields
\begin{eqnarray*}
  \mathbb{P}_{t_{\timeInd+1}}(\mathbf{n})
&=&
\sum_{\mathbf{m}} \mathbb{P}_{t_{\timeInd}}(\mathbf{m}) \frac{\tilde{\gamma}^\timeInd_{\mathbf{m}}}{\tilde{\gamma}^\timeInd_{\mathbf{n}}}
\mathbb{P}(C_{t_{\timeInd+1}} = n^{(c)}
\mid C_{t_{\timeInd}} = m^{(c)})
\\ && \qquad \times \qquad
\mathbb{P}(\moranModified_{t_{\timeInd+1}} = \mathbf{n}
\mid C_{t_{\timeInd+1}}= n^{(c)}, C_{t_{\timeInd}} = m^{(c)}, \moranModified_{t_{\timeInd}} = \mathbf{m})
\\
&=&
\sum_{\mathbf{m}} \mathbb{P}_{t_{\timeInd}}(\mathbf{m}) \frac{\tilde{\gamma}^\timeInd_{\mathbf{m}}}{\tilde{\gamma}^\timeInd_{\mathbf{n}}}
\mathbb{P}(C_{t_{\timeInd+1}}= n^{(c)}, \moranModified_{t_{\timeInd+1}} = \mathbf{n}
\mid C_{t_{\timeInd}} = m^{(c)}, \moranModified_{t_{\timeInd}} = \mathbf{m})
\\
&=&
\sum_{\mathbf{m}} \mathbb{P}_{t_{\timeInd}}(\mathbf{m}) \frac{\tilde{\gamma}^\timeInd_{\mathbf{m}}}{\tilde{\gamma}^\timeInd_{\mathbf{n}}} \Big[e^{\tilde{\Rate}^\timeInd (t_{\timeInd+1} - t_{\timeInd})}\Big]_{\mathbf{m}, \mathbf{n}}.
\end{eqnarray*}
which proves half of the desired result, i.e.
$\bfp^{\timeInd+1} = \big[(\bfp^{\timeInd} \odot \tilde{\bfmath{\gamma}}^{\timeInd})
e^{\tilde{\Lambda}^\timeInd (t_{\timeInd+1} - t_\timeInd)}\big] \div \tilde{\bfmath{\gamma}}^{\timeInd}$,
where $\bfp^\timeInd = [\mathbb{P}_{t_\timeInd}(\mathbf{n})]_{\mathbf{n}}'$.
To show the other half, that
$  \bfp^{-\pieces+1} = \tilde{\bfmath{\lambda}}^{-\pieces} \div \tilde{\bfmath{\gamma}}^{-\pieces}$,
where $\tilde{\bfmath{\lambda}}^{-\pieces}$ is the stationary distribution of $\Rate^{-\pieces}$,
we simply note that
for all $t \leq t_{-\pieces+1}$,
\begin{eqnarray*}
  \mathbb{P}_t(\mathbf{n}) \tilde{\gamma}^{-\pieces}_{\mathbf{n}}
&=&
  \mathbb{P}(\moranModified^*_t = \mathbf{n} \mid C^*_t = n^{(c)}) \gamma^{-\pieces}_{n^{(c)}}
\\
&=&
  \mathbb{P}(\moranModified_t = \mathbf{n} \mid C_t = n^{(c)}) \gamma^{-\pieces}_{n^{(c)}}
\\
&=&
  \mathbb{P}(\moranModified_t = \mathbf{n})
\\
&=& \tilde{\lambda}^{-\pieces}_{\mathbf{n}},
\end{eqnarray*}
where the second equality follows by reversibility
of $\bfGamma^{-\pieces}$,
which implies
$\mathbb{P}(C_{\leq t} \mid C_t) = \mathbb{P}(C^*_{\leq t} \mid C^*_t)$,
and thus $\mathbb{P}(\moranModified_{\leq t} \mid C_t) = \mathbb{P}(\moranModified^*_{\leq t} \mid C^*_t)$.

\subsection{Proof of Theorem~\ref{thm:posterior:chain} \label{proof:posterior:chain}}

We first check that $\mathbb{P}(\mathbf{n}_{s_1} \mid
\mathbf{n}_{s_2}, \mathbf{n}_{s_3}) = \mathbb{P}(\mathbf{n}_{s_1} \mid \mathbf{n}_{s_2})$,
for $-\infty < s_1 < s_2 < s_3 \leq 0$,
and so $\mathbf{n}_t$ is a backwards in time Markov chain.

\begin{figure}
  \centering
  \includegraphics[width=0.9\textwidth]{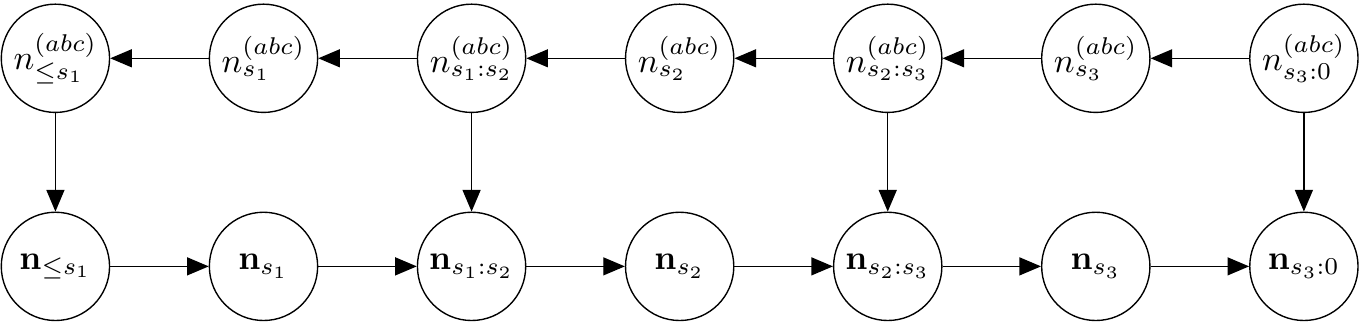}
  \caption{Probabilistic graphical model for the coalescent with recombination
    and mutation, with $-\infty < s_1 < s_2 < s_3 \leq
    0$.}
  \label{fig:directed:graphical:model}
\end{figure}

Recall that we generate $n^{(abc)}_t$ as a backwards in
time Markov chain, then generate $\mathbf{n}_t$ by dropping down
mutations forward in time.
The conditional independence structure of $\mathbf{n}_{s_1},
\mathbf{n}_{s_2}, \mathbf{n}_{s_3}$ is thus described by the
directed graphical model \citep{koller2009probabilistic} in
Figure~\ref{fig:directed:graphical:model}.

\begin{figure}
 \centering
  \includegraphics[width=0.33\textwidth]{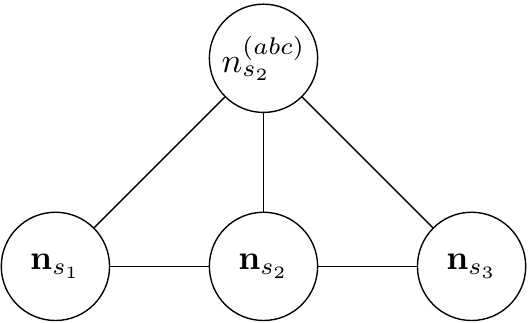}
  \caption{Undirected graphical model, after moralization and
    variable elimination on Figure~\ref{fig:directed:graphical:model}: we
    add edges to form cliques on the left and right sides of
    $n^{(abc)}_{s_2}, \mathbf{n}_{s_2}$, and then eliminate
    all the variables except the ones pictured here.}
  \label{fig:undirected:graphical:model}
\end{figure}

Doing moralization and variable elimination \citep{koller2009probabilistic} on Figure~\ref{fig:directed:graphical:model} results
in the undirected graphical model in
Figure~\ref{fig:undirected:graphical:model}.
The graphical model of Figure~\ref{fig:undirected:graphical:model}
then implies
\begin{eqnarray*}
  \mathbb{P}(\mathbf{n}_{s_1} \mid \mathbf{n}_{s_2}, \mathbf{n}_{s_3})
&=&
\sum_{n^{(abc)}_{s_2}}
\mathbb{P}(\mathbf{n}_{s_1} \mid \mathbf{n}_{s_2}, n^{(abc)}_{s_2})
\mathbb{P}(n^{(abc)}_{s_2} \mid \mathbf{n}_{s_2}, \mathbf{n}_{s_3})
\\
&=&
\mathbb{P}(\mathbf{n}_{s_1} \mid \mathbf{n}_{s_2}),
\end{eqnarray*}
where the second equality follows because $n^{(abc)}_{s_2}$ is a deterministic function of
$\mathbf{n}_{s_2}$.
Thus, $\mathbf{n}_t$ is a backwards in time Markov chain.

We next compute the backwards in time rates
$\isRate^{(t)}_{\mathbf{n},\mathbf{m}}$ for the Markov chain
$\mathbf{n}_t$ at time $t$. Starting from the definition of $\isRate^{(t)}_{\mathbf{n},\mathbf{m}}$,
\begin{eqnarray*}
  \isRate^{(t)}_{\mathbf{n},\mathbf{m}}
&=&
\frac{d}{ds} \mathbb{P}(\mathbf{n}_{t-s} = \mathbf{m} \mid
\mathbf{n}_{t} = \mathbf{n}) \Big|_{s=0}
\\
&=&
\frac{d}{ds} \frac{\mathbb{P}(\mathbf{n}_{t-s} = \mathbf{m} ,
\mathbf{n}_{t} = \mathbf{n} \mid n^{(abc)}_t =
n^{(abc)})}{\mathbb{P}(\mathbf{n}_t = \mathbf{n} \mid n^{(abc)}_t = n^{(abc)})} \Big|_{s=0}
\\
&=&
\frac1{\mathbb{P}_t(\mathbf{n})}
\frac{d}{ds} 
\Big[
\mathbb{P}(n^{(abc)}_{t-s} = m^{(abc)} \mid n^{(abc)}_{t} =  n^{(abc)})
\mathbb{P}(\mathbf{n}_{t} = \mathbf{n} \mid n^{(abc)}_{t} = n^{(abc)},\mathbf{n}_{t-s} = \mathbf{m})
\mathbb{P}_{t-s}(\mathbf{m})
\Big]\Big|_{s=0}
\\
&=&
\frac1{\mathbb{P}_t(\mathbf{n})}
\Big[
\mathbb{P}(n^{(abc)}_{t} = m^{(abc)} \mid n^{(abc)}_{t} =
n^{(abc)})
\mathbb{P}(\mathbf{n}_{t} = \mathbf{n} \mid n^{(abc)}_{t} =
n^{(abc)},\mathbf{n}_{t} = \mathbf{m})  \frac{d}{ds} \mathbb{P}_{t-s}(\mathbf{m}) \Big|_{s=0} 
\\
&&  \qquad \qquad 
+
\isQ^{(t)}_{\mathbf{n},\mathbf{m}}
\mathbb{P}_t(\mathbf{m})
\Big]
\\
&=&
\begin{cases}
  \isQ^{(t)}_{\mathbf{n},\mathbf{m}}
\frac{\mathbb{P}_t(\mathbf{m})}{\mathbb{P}_t(\mathbf{n})}, & \text{if }
\mathbf{m} \neq \mathbf{n}, \\
\isQ^{(t)}_{\mathbf{n},\mathbf{n}} - \frac{d}{dt} \log
\mathbb{P}_t(\mathbf{n}),
& \text{if } \mathbf{m} = \mathbf{n},
\end{cases}
\end{eqnarray*}
where the penultimate equality follows from the product rule and the
definition of $\isQbold^{(t)}$ in \eqref{eq:Q}.

The specific entries of $\isQbold^{(t)}$ listed in Table~\ref{tab:Q} can be
obtained by applying the product rule to \eqref{eq:Q}, and noting that $\frac{d}{ds}
\mathbb{P}(n_{t-s}^{(abc)} \mid n_t^{(abc)})|_{s=0}$ and $\frac{d}{ds}
\mathbb{P}(\mathbf{n}_t \mid n_t^{(abc)}, \mathbf{n}_{t-s}) |_{s=0}$
are, respectively, the backwards in time rates of $n^{(abc)}_t$ (as listed in
Table~\ref{tab:arg:rates}), and the forward in time rates for dropping
mutations on $\mathbf{n}_t$.

\clearpage
\bibliographystyle{myplainnat}  
\bibliography{bib/yss-group}

\providecommand{\sortkey}[1]{}
\begin{thebibliography}{59}
\providecommand{\natexlab}[1]{#1}
\providecommand{\url}[1]{\texttt{#1}}
\expandafter\ifx\csname urlstyle\endcsname\relax
  \providecommand{\doi}[1]{doi: #1}\else
  \providecommand{\doi}{doi: \begingroup \urlstyle{rm}\Url}\fi

\bibitem[{1000 Genomes Project Consortium}(2010)]{gpc2010map}
{1000 Genomes Project Consortium}.
\newblock 2010.
\newblock A map of human genome variation from population-scale sequencing.
\newblock \emph{Nature}, {\bf 467,} 1061--1073.

\bibitem[Al-Mohy and Higham(2011)]{al-mohy2011computing}
Al-Mohy, A.~H. and Higham, N.~J.
\newblock 2011.
\newblock Computing the action of the matrix exponential, with an application
  to exponential integrators.
\newblock \emph{SIAM Journal on Scientific Computing}, {\bf 33 (2),} 488--511.

\bibitem[Auton and McVean(2007)]{auton2007recombination}
Auton, A. and McVean, G.
\newblock 2007.
\newblock Recombination rate estimation in the presence of hotspots.
\newblock \emph{Genome Research}, {\bf 17,}\penalty0 (8) 1219--1227.

\bibitem[Auton et~al.(2012)Auton, Fledel-Alon, Pfeifer, Venn, S{\'e}gurel,
  Street, Leffler, Bowden, Aneas, Broxholme, et~al.]{auton2012fine}
Auton, A., Fledel-Alon, A., Pfeifer, S., Venn, O., S{\'e}gurel, L., Street, T.,
  Leffler, E.~M., Bowden, R., Aneas, I., Broxholme, J., et~al.
\newblock 2012.
\newblock A fine-scale chimpanzee genetic map from population sequencing.
\newblock \emph{Science}, {\bf 336,}\penalty0 (6078) 193--198.

\bibitem[Auton et~al.(2013)Auton, Li, Kidd, Oliveira, Nadel, Holloway, Hayward,
  Cohen, Greally, Wang, et~al.]{auton2013genetic}
Auton, A., Li, Y.~R., Kidd, J., Oliveira, K., Nadel, J., Holloway, J.~K.,
  Hayward, J.~J., Cohen, P.~E., Greally, J.~M., Wang, J., et~al.
\newblock 2013.
\newblock Genetic recombination is targeted towards gene promoter regions in
  dogs.
\newblock \emph{{PLoS Genetics}}, {\bf 9,}\penalty0 (12) e1003984.

\bibitem[Auton et~al.(2014)Auton, Myers, and McVean]{auton2014identifying}
Auton, A., Myers, S., and McVean, G.
\newblock Identifying recombination hotspots using population genetic data.
\newblock arXiv preprint: http://arxiv.org/abs/1403.4264, March 2014.

\bibitem[Baudat et~al.(2010)Baudat, Buard, Grey, Fledel-Alon, Ober, Przeworski,
  Coop, and de~Massy]{baudat2010prdm9}
Baudat, F., Buard, J., Grey, C., Fledel-Alon, A., Ober, C., Przeworski, M.,
  Coop, G., and de~Massy, B.
\newblock 2010.
\newblock {PRDM9} is a major determinant of meiotic recombination hotspots in
  humans and mice.
\newblock \emph{Science}, {\bf 327,} 836--840.

\bibitem[Berg et~al.(2010)Berg, Neumann, Lam, Sarbajna, Odenthal-Hesse, May,
  and Jeffreys]{berg2010prdm9}
Berg, I.~L., Neumann, R., Lam, K.~G., Sarbajna, S., Odenthal-Hesse, L., May,
  C.~A., and Jeffreys, A.~J.
\newblock 2010.
\newblock {PRDM9} variation strongly influences recombination hot-spot activity
  and meiotic instability in humans.
\newblock \emph{Nature Genetics}, {\bf 42,}\penalty0 (10) 859--863.

\bibitem[Bhaskar and Song(2012)]{bhaskar2012closed}
Bhaskar, A. and Song, Y.~S.
\newblock 2012.
\newblock Closed-form asymptotic sampling distributions under the coalescent
  with recombination for an arbitrary number of loci.
\newblock \emph{Advances in Applied Probability}, {\bf 44,} 391--407.
\newblock (PMC3409093).

\bibitem[Chan et~al.(2012)Chan, Jenkins, and Song]{chan2012genome}
Chan, A.~H., Jenkins, P.~A., and Song, Y.~S.
\newblock 2012.
\newblock Genome-wide fine-scale recombination rate variation in {D}rosophila
  melanogaster.
\newblock \emph{{PLoS Genetics}}, {\bf 8,}\penalty0 (12) e1003090.

\bibitem[Chen et~al.(2009)Chen, Marjoram, and Wall]{chen2009fast}
Chen, G.~K., Marjoram, P., and Wall, J.~D.
\newblock 2009.
\newblock Fast and flexible simulation of {DNA} sequence data.
\newblock \emph{Genome Res.}, {\bf 19,} 136--142.

\bibitem[Choudhary and Singh(1987)]{choudhary1987historical}
Choudhary, M. and Singh, R.
\newblock 1987.
\newblock Historical effective size and the level of genetic diversity in
  drosophila melanogaster and drosophila pseudoobscura.
\newblock \emph{Biochemical genetics}, {\bf 25,}\penalty0 (1-2) 41--51.

\bibitem[{De~Iorio} and Griffiths(2004)]{deiorio2004importance}
{De~Iorio}, M. and Griffiths, R.~C.
\newblock 2004.
\newblock Importance sampling on coalescent histories. {I}.
\newblock \emph{Adv. Appl. Prob.}, {\bf 36,} 417--433.

\bibitem[Dialdestoro et~al.(2016)Dialdestoro, Sibbesen, Maretty, Raghwani,
  Gall, Kellam, Pybus, Hein, and Jenkins]{dialdestoro2016coalescent}
Dialdestoro, K., Sibbesen, J.~A., Maretty, L., Raghwani, J., Gall, A., Kellam,
  P., Pybus, O.~G., Hein, J., and Jenkins, P.~A.
\newblock 2016.
\newblock Coalescent inference using serially sampled, high-throughput
  sequencing data from intra-host hiv infection.
\newblock \emph{Genetics}.
\newblock ISSN 0016-6731.
\newblock \doi{10.1534/genetics.115.177931}.
\newblock URL
  \url{http://www.genetics.org/content/early/2016/02/03/genetics.115.177931}.

\bibitem[Donnelly and Kurtz(1999)]{donnelly1999genealogical}
Donnelly, P. and Kurtz, T.~G.
\newblock 1999.
\newblock Genealogical processes for fleming-viot models with selection and
  recombination.
\newblock \emph{Annals of Applied Probability}, {\bf 9,}\penalty0 (4)
  1091--1148.

\bibitem[Durrett(2008)]{durrett2008probability}
Durrett, R.
\newblock \emph{Probability Models for DNA Sequence Evolution}.
\newblock Springer, New York, 2nd edition, 2008.

\bibitem[Ethier and Griffiths(1990)]{ethier1990two}
Ethier, S.~N. and Griffiths, R.~C.
\newblock 1990.
\newblock On the two-locus sampling distribution.
\newblock \emph{J. Math. Biol.}, {\bf 29,} 131--159.

\bibitem[Ethier and Kurtz(1993)]{ethier1993fleming}
Ethier, S.~N. and Kurtz, T.~G.
\newblock 1993.
\newblock Fleming-viot processes in population genetics.
\newblock \emph{SIAM Journal on Control and Optimization}, {\bf 31,}\penalty0
  (2) 345--386.

\bibitem[Fearnhead(2003)]{fearnhead2003consistency}
Fearnhead, P.
\newblock 2003.
\newblock Consistency of estimators of the population-scaled recombination
  rate.
\newblock \emph{Theoretical Population Biology}, {\bf 64,} 67--79.

\bibitem[Fearnhead and Donnelly(2001)]{fearnhead2001estimating}
Fearnhead, P. and Donnelly, P.
\newblock 2001.
\newblock Estimating recombination rates from population genetic data.
\newblock \emph{Genetics}, {\bf 159,} 1299--1318.

\bibitem[Fearnhead and Smith(2005)]{fearnhead2005novel}
Fearnhead, P. and Smith, N. G.~C.
\newblock 2005.
\newblock A novel method with improved power to detect recombination hotspots
  from polymorphism data reveals multiple hotspots in human genes.
\newblock \emph{Am. J. Hum. Genet.}, {\bf 77,} 781--794.

\bibitem[Fearnhead et~al.(2004)Fearnhead, Harding, Schneider, Myers, and
  Donnelly]{fearnhead2004application}
Fearnhead, P., Harding, R.~M., Schneider, J.~A., Myers, S., and Donnelly, P.
\newblock 2004.
\newblock Application of coalescent methods to reveal fine-scale rate variation
  and recombination hotspots.
\newblock \emph{Genetics}, {\bf 167,} 2067--2081.

\bibitem[Fearnhead(2006)]{fearnhead2006sequenceldhot}
Fearnhead, P.
\newblock 2006.
\newblock Sequence{LD}hot: detecting recombination hotspots.
\newblock \emph{Bioinformatics}, {\bf 22,} 3061--3066.

\bibitem[Golding(1984)]{golding1984sampling}
Golding, G.~B.
\newblock 1984.
\newblock The sampling distribution of linkage disequilibrium.
\newblock \emph{Genetics}, {\bf 108,} 257--274.

\bibitem[Griffiths and Marjoram(1997)]{griffiths1997ancestral}
Griffiths, R.~C. and Marjoram, P.
\newblock An ancestral recombination graph.
\newblock In Donnelly, P. and Tavar\'e, S., editors, \emph{Progress in
  population genetics and human evolution}, volume~87, pages 257--270.
  Springer-Verlag, Berlin, 1997.

\bibitem[Griffiths et~al.(2008)Griffiths, Jenkins, and
  Song]{griffiths2008importance}
Griffiths, R.~C., Jenkins, P.~A., and Song, Y.~S.
\newblock 2008.
\newblock Importance sampling and the two-locus model with subdivided
  population structure.
\newblock \emph{Advances in Applied Probability}, {\bf 40,} 473--500.

\bibitem[Griffiths(1991)]{griffiths1991two}
Griffiths, R.
\newblock 1991.
\newblock The two-locus ancestral graph.
\newblock \emph{Selected Proceedings of the Sheffield Symposium on Applied
  Probability. IMS Lecture Notes--Monograph Series}, {\bf 18,} 100--117.

\bibitem[Gutenkunst et~al.(2009)Gutenkunst, Hernandez, Williamson, and
  Bustamante]{gutenkunst2009inferring}
Gutenkunst, R.~N., Hernandez, R.~D., Williamson, S.~H., and Bustamante, C.~D.
\newblock 2009.
\newblock Inferring the joint demographic history of multiple populations from
  multidimensional {SNP} frequency data.
\newblock \emph{PLoS Genetics}, {\bf 5,}\penalty0 (10) e1000695.

\bibitem[Hobolth et~al.(2008)Hobolth, Uyenoyama, and
  Wiuf]{hobolth2008importance}
Hobolth, A., Uyenoyama, M., and Wiuf, C.
\newblock 2008.
\newblock Importance sampling for the infinite sites model.
\newblock \emph{Statistical Applications in Genetics and Molecular Biology},
  {\bf 7,}\penalty0 (1) 32.

\bibitem[Hudson(1985)]{hudson1985sampling}
Hudson, R.~R.
\newblock 1985.
\newblock Sampling distribution of linkage disequilibrium under an infinite
  allele model without selection.
\newblock \emph{Genetics}, {\bf 109,} 611--631.

\bibitem[Hudson(2001)]{hudson2001two}
Hudson, R.
\newblock 2001.
\newblock Two-locus sampling distributions and their application.
\newblock \emph{Genetics}, {\bf 159,}\penalty0 (4) 1805--1817.

\bibitem[Jenkins and Song(2009)]{jenkins2009closed}
Jenkins, P.~A. and Song, Y.~S.
\newblock 2009.
\newblock Closed-form two-locus sampling distributions: accuracy and
  universality.
\newblock \emph{Genetics}, {\bf 183,} 1087--1103.

\bibitem[Jenkins and Song(2010)]{jenkins2010asymptotic}
Jenkins, P.~A. and Song, Y.~S.
\newblock 2010.
\newblock An asymptotic sampling formula for the coalescent with recombination.
\newblock \emph{Annals of Applied Probability}, {\bf 20,} 1005--1028.
\newblock (PMC2910927).

\bibitem[Jenkins(2012)]{jenkins2012stopping}
Jenkins, P.
\newblock 2012.
\newblock Stopping-time resampling and population genetic inference under
  coalescent models.
\newblock \emph{Statistical Applications in Genetics and Molecular Biology},
  {\bf 11,}\penalty0 (1) 1--20.
\newblock (PMC3800802).

\bibitem[Jenkins and Song(2012)]{jenkins2012pade}
Jenkins, P.~A. and Song, Y.~S.
\newblock 2012.
\newblock Pad\'e approximants and exact two-locus sampling distributions.
\newblock \emph{Annals of Applied Probability}, {\bf 22,} 576--607.
\newblock (PMC3685441).

\bibitem[Johnson and Slatkin(2009)]{johnson2009inference}
Johnson, P. and Slatkin, M.
\newblock 2009.
\newblock Inference of microbial recombination rates from metagenomic data.
\newblock \emph{PLoS Genetics}, {\bf 5,}\penalty0 (10) e1000674.

\bibitem[Johnston and Cutler(2012)]{johnston2012population}
Johnston, H.~R. and Cutler, D.~J.
\newblock 2012.
\newblock Population demographic history can cause the appearance of
  recombination hotspots.
\newblock \emph{The American Journal of Human Genetics}, {\bf 90,}\penalty0 (5)
  774--783.

\bibitem[Kamm et~al.(2016)Kamm, Terhorst, and Song]{kamm2016efficient}
Kamm, J.~A., Terhorst, J., and Song, Y.~S.
\newblock 2016.
\newblock Efficient computation of the joint sample frequency spectra for
  multiple populations.
\newblock \emph{Journal of Computational and Graphical Statistics}, pages
  1--37.
\newblock \doi{10.1080/10618600.2016.1159212}.
\newblock URL \url{http://dx.doi.org/10.1080/10618600.2016.1159212}.

\bibitem[Koller and Friedman(2009)]{koller2009probabilistic}
Koller, D. and Friedman, N.
\newblock \emph{Probabilistic graphical models: principles and techniques}.
\newblock MIT press, 2009.

\bibitem[Koskela et~al.(2015)Koskela, Jenkins, and
  Spano]{koskela2015computational}
Koskela, J., Jenkins, P.~A., and Spano, D.
\newblock 2015.
\newblock Computational inference beyond kingman's coalescent.
\newblock \emph{Journal of Applied Probability}, {\bf 52,}\penalty0 (2)
  519--537.

\bibitem[Maruyama(1982)]{maruyama1982stochastic}
Maruyama, T.
\newblock Stochastic integrals and their application to population genetics.
\newblock In Kimura, M., editor, \emph{Molecular Evolution, Protein
  Polymorphism and their Neutral Theory}, pages 151--166. Springer-Verlag,
  Berlin, 1982.

\bibitem[McVean et~al.(2002)McVean, Awadalla, and
  Fearnhead]{mcvean2002coalescent}
McVean, G., Awadalla, P., and Fearnhead, P.
\newblock 2002.
\newblock A coalescent-based method for detecting and estimating recombination
  from gene sequences.
\newblock \emph{Genetics}, {\bf 160,} 1231--1241.

\bibitem[Mc{V}ean(2002)]{mcvean2002genealogical}
Mc{V}ean, G. A.~T.
\newblock 2002.
\newblock A genealogical interpretation of linkage disequilibrium.
\newblock \emph{Genetics}, {\bf 162,} 987--991.

\bibitem[McVean et~al.(2004)McVean, Myers, Hunt, Deloukas, Bentley, and
  Donnelly]{mcvean2004fine}
McVean, G., Myers, S., Hunt, S., Deloukas, P., Bentley, D., and Donnelly, P.
\newblock 2004.
\newblock The fine-scale structure of recombination rate variation in the human
  genome.
\newblock \emph{Science}, {\bf 304,}\penalty0 (5670) 581--584.

\bibitem[Moran(1958)]{moran1958random}
Moran, P.
\newblock 1 1958.
\newblock Random processes in genetics.
\newblock \emph{Mathematical Proceedings of the Cambridge Philosophical
  Society}, {\bf 54,} 60--71.
\newblock ISSN 1469-8064.
\newblock \doi{10.1017/S0305004100033193}.

\bibitem[Myers et~al.(2005)Myers, Bottolo, Freeman, McVean, and
  Donnelly]{myers2005fine}
Myers, S., Bottolo, L., Freeman, C., McVean, G., and Donnelly, P.
\newblock 2005.
\newblock A fine-scale map of recombination rates and hotspots across the human
  genome.
\newblock \emph{Science}, {\bf 310,}\penalty0 (5746) 321--324.

\bibitem[Myers et~al.(2010)Myers, Bowden, Tumian, Bontrop, Freeman, MacFie,
  McVean, and Donnelly]{myers2010drive}
Myers, S., Bowden, R., Tumian, A., Bontrop, R., Freeman, C., MacFie, T.,
  McVean, G., and Donnelly, P.
\newblock 2010.
\newblock Drive against hotspot motifs in primates implicates the {PRDM9} gene
  in meiotic recombination.
\newblock \emph{Science}, {\bf 327,}\penalty0 (5967) 876--879.

\bibitem[Myers et~al.(2008)Myers, Freeman, Auton, Donnelly, and
  McVean]{myers2008common}
Myers, S., Freeman, C., Auton, A., Donnelly, P., and McVean, G.
\newblock 2008.
\newblock A common sequence motif associated with recombination hot spots and
  genome instability in humans.
\newblock \emph{Nature Genetics}, {\bf 40,}\penalty0 (9) 1124--1129.

\bibitem[Ohta and Kimura(1969)]{ohta1969linkage}
Ohta, T. and Kimura, M.
\newblock 1969.
\newblock Linkage disequilibrium due to random genetic drift.
\newblock \emph{Genet. Res. Camb.}, {\bf 13,} 47--55.

\bibitem[Sheehan et~al.(2013)Sheehan, Harris, and Song]{sheehan2013estimating}
Sheehan, S., Harris, K., and Song, Y.~S.
\newblock 2013.
\newblock Estimating variable effective population sizes from multiple genomes:
  A sequentially {M}arkov conditional sampling distribution approach.
\newblock \emph{Genetics}, {\bf 194,}\penalty0 (3) 647--662.
\newblock (PMC3697970).

\bibitem[Smith and Fearnhead(2005)]{smith2005comparison}
Smith, N. G.~C. and Fearnhead, P.
\newblock 2005.
\newblock A comparison of three estimators of the population-scaled
  recombination rate: Accuracy and robustness.
\newblock \emph{Genetics}, {\bf 171,}\penalty0 (4) 2051--2062.
\newblock ISSN 0016-6731.
\newblock \doi{10.1534/genetics.104.036293}.
\newblock URL \url{http://www.genetics.org/content/171/4/2051}.

\bibitem[Song and Song(2007)]{song2007analytic}
Song, Y.~S. and Song, J.~S.
\newblock 2007.
\newblock Analytic computation of the expectation of the linkage disequilibrium
  coefficient $r^2$.
\newblock \emph{Theoretical Population Biology}, {\bf 71,} 49--60.

\bibitem[Steinr{\"u}cken et~al.(2015)Steinr{\"u}cken, Kamm, and
  Song]{steinruecken2015inference}
Steinr{\"u}cken, M., Kamm, J.~A., and Song, Y.~S.
\newblock Inference of complex population histories using whole-genome
  sequences from multiple populations.
\newblock bioRxiv preprint: http://dx.doi.org/10.1101/026591, September 2015.

\bibitem[Stephens and Donnelly(2000)]{stephens2000inference}
Stephens, M. and Donnelly, P.
\newblock 2000.
\newblock Inference in molecular population genetics.
\newblock \emph{J.R. Stat. Soc. Ser. B}, {\bf 62,} 605--655.

\bibitem[Tajima(1983)]{tajima1983evolutionary}
Tajima, F.
\newblock 1983.
\newblock Evolutionary relationship of {DNA} sequences in finite populations.
\newblock \emph{Genetics}, {\bf 105,}\penalty0 (2) 437--460.

\bibitem[{The International HapMap Consortium}(2007)]{ihmc2007seconda}
{The International HapMap Consortium}.
\newblock 2007.
\newblock A second generation human haplotype map of over 3.1 million {SNPs}.
\newblock \emph{Nature}, {\bf 449,}\penalty0 (7164) 851--861.

\bibitem[Wegmann et~al.(2011)Wegmann, Kessner, Veeramah, Mathias, Nicolae,
  Yanek, Sun, Torgerson, Rafaels, Mosley, Becker, Ruczinski, Beaty, Kardia,
  Meyers, Barnes, Becker, Freimer, and Novembre]{wegmann2011recombination}
Wegmann, D., Kessner, D.~E., Veeramah, K.~R., Mathias, R.~A., Nicolae, D.~L.,
  Yanek, L.~R., Sun, Y.~V., Torgerson, D.~G., Rafaels, N., Mosley, T., Becker,
  L.~C., Ruczinski, I., Beaty, T.~H., Kardia, S. L.~R., Meyers, D.~A., Barnes,
  K.~C., Becker, D.~M., Freimer, N.~B., and Novembre, J.
\newblock 2011.
\newblock Recombination rates in admixed individuals identified by
  ancestry-based inference.
\newblock \emph{Nat. Genet.}, {\bf 43,} 847--853.

\bibitem[Weir(1996)]{weir1996genetic}
Weir, B.
\newblock \emph{Genetic data analysis II: Methods for discrete population
  genetic data.}
\newblock Sinauer Associates, Sunderland, MA, 1996.

\bibitem[Ye et~al.(2013)Ye, Nielsen, Nicholson, Teh, Jenkins, Colchester,
  Anderson, and Hein]{ye2013importance}
Ye, M., Nielsen, S., Nicholson, M., Teh, Y., Jenkins, P., Colchester, F.,
  Anderson, J., and Hein, J.
\newblock Importance sampling under the coalescent with times and variable
  population.
\newblock URL
  \url{https://www.stats.ox.ac.uk/__data/assets/pdf_file/0007/9889/Coalescent_Sampling_Report.pdf}.
\newblock 2013.

\end{thebibliography}
\end{document}